\documentclass{article}

\usepackage{arxiv}

\usepackage[utf8]{inputenc} 
\usepackage[T1]{fontenc}    
\usepackage{hyperref}       
\usepackage{url}            
\usepackage{booktabs}       
\usepackage{amsfonts}       
\usepackage{nicefrac}       
\usepackage{microtype}      
\usepackage{lipsum}
\usepackage{amsmath}
\usepackage{array}
\usepackage{pict2e}
\usepackage{slashbox}
\usepackage{caption}
\usepackage{subcaption}
\usepackage{tikz}
\usetikzlibrary{quantikz}

\usepackage{hyperref}

\usepackage[
backend=biber,
citestyle=numeric
]{biblatex}
\title{Laplacian Eigenmaps with variational circuits: a quantum embedding of graph data. }

\author{
  Slimane Thabet \\
  EDF R\&D UK Center, Hove, United Kingdom\\
  Ecole polytechnique, Palaiseau, France\\
  \texttt{slimane.thabet@polytechnique.edu} \\
   \And
 Jean-François Hullo \\
  EDF R\&D UK Center, Hove, United Kingdom\\
  \texttt{Jean-Francois.Hullo@edf.fr} \\
}

\begin{document}
\maketitle

\begin{abstract}
With the development of quantum algorithms, high-cost computations are being scrutinized in the hope of a quantum advantage. While graphs offer a convenient framework for multiple real-world problems, their analytics still comes with high computation and space. By mapping the graph data into a low dimensional space, in which graph structural information is preserved, the eigenvectors of the Laplacian matrix constitute a powerful node embedding, called Laplacian Eigenmaps. Computing these embeddings is on its own an expensive task knowing that using specific sparse methods, the eigendecomposition of a Laplacian matrix has a cost of O($rn^2$), $r$ being the ratio of nonzero elements.

We propose a method to compute a Laplacian Eigenmap using a quantum variational circuit. The idea of our algorithm is to reach the eigenstates of the laplacian matrix, which can be considered as a hamiltonian operator, by adapting the variational quantum eigensolver algorithm. By estimating the $d$ first eigenvectors of the Laplacian at the same time, our algorithm directly generates a $d$ dimension quantum embedding of the graph. We demonstrate that it is possible to use the embedding for graph machine learning tasks by implementing a quantum classifier on the top of it. The overall circuit consists in a full quantum node classification algorithm. Tests on 32 nodes graph with a quantum simulator shows that we can achieve similar performances as the classical laplacian eigenmap algorithm. Although mathematical properties of this approximate approach are not fully understood, this algorithm opens perspectives for graph pre-processing using noisy quantum computers.
\end{abstract}


\tableofcontents

\section{Introduction}

A graph is a versatile mathematical object that can be used for modelling a lot of real world objects. It is a set of vertices linked together by a set of edges which makes it suitable to represent many kinds of connected objects. Adding weights to the edges or the nodes allows to express the differences of strength between the bonds.

Many natural networks exist, such as social networks, computer networks, molecules. Networks can also be created from other contexts, such as word relationships, when two words are connected if they appear in the same sentence. Graphs are also very useful for modelling several optimization problems \cite{commander2008, laporte1990selective}. 

Many graph problems can be solved with graph embeddings \cite{Goyal2018}, especially machine learning ones. An embedding in the general case is a map from a metric space to an euclidean space such as the distance between the images are close to the distance between the objects. Embeddings are very used in Natural language Processing, where word embeddings are scrutinized \cite{Bengio2003}. A word embedding is a vector representation of words such as the distance between two vector images is representative of the semantic similarity.

Similarly to word embeddings, a graph embedding is a map from the set of nodes of a particular graph to an euclidean space such as the distances between the images reflect the similarity between the nodes in the graph. The aim is to obtain a set of vectors which captures structural patterns of the graph, for example communities. We would have that nodes belonging to the same community will be closely embedded.

Graph embeddings and embeddings in general are widely used for machine learning tasks. Most of machine learning algorithms take vectors as input, so computing an embedding is a necessary preprocessing step before applying common algorithms such as logistic regression, SVM, or neural networks.

Computing a graph embedding is not a simple task, and can become computationally expensive. Lots of methods exist, which can capture different properties of the graph. \cite{Goyal2018} made an exhaustive survey of the main methods and compared their performance on standards tasks. One of the most simple method to compute a $d$ dimension graph embedding is to take the eigenvectors of the laplacian matrix corresponding to the $d$ smallest eigenvalues \cite{Belkin2002}. It is called Laplacian Eigenmap and involves finding eigenvectors of a real symmetric matrix. Clustering and classification algorithm can then be applied to this embedding for machine learning purposes. Applying a clustering algorithm  such as k-means on the top of Laplacian eigenmap is sometimes called spectral clustering \cite{VonLuxburg2007}. The goal of this paper is to explore a way to construct a graph embedding with a quantum computer, in a way that looks like Laplacian eigenmaps.

Quantum computing is the use of quantum properties of the matter to create a new computing paradigm. Instead of using classical bits, which have a deterministic value 0 or 1, quantum computers use quantum bits or qubits, which are in a probabilistic state between 0 or 1 \cite{M.A.NielsenandI.L.Chuang2011}. Once measured, it collapses to the state 0 or the state 1. A quantum state is the state of a set of qubits. It can then be considered as a probability distribution, and measuring it is sampling from the underlying distribution.

The theoretical promises of quantum computing are tremendous. In 1998, Shor elaborated a quantum algorithm that could factorize numbers with an exponential speedup \cite{shor}. Grover showed that a seach in an unstructured database of $N$ elements can be performed in $\mathcal{O}(\sqrt{N})$ time \cite{grover1996fast}. Quantum computing could then reduce the time to perform highly complex and costly operations. Applications in which there is an active research for quantum advantage are chemistry, optimization, and machine learning. 

Unfortunately, the current hardware only allows to execute a circuit of maximum a few dozen of qubits and a few dozen of gates \cite{pino2020demonstration} where millions are needed for practical problems. Today, quantum hardware is in what is called the Noisy Intermediate Scale Quantum Computing (NISQ) era \cite{Preskill2018quantumcomputingin}. However, quantum hardware makes continuous improvement both in fidelity and number of qubits, and in 2019 Google announced to have reached quantum supremacy \cite{Arute2019}. It means that for the first time, a quantum computer was able to execute a task faster than any classical computer. This achievement encourages to continue the research on quantum algorithms for many types of problems.

Quantum machine learning is the intersection between quantum computing and machine learning \cite{Schuld2018}. Machine learning is creating a model of the world from observed data. Then quantum machine learning can be understood in 3 ways: applying classical models to quantum data, quantum models to classical data, and quantum models to quantum data. Most of research is done on the second category, using quantum computers to create models for classical data.

The goal of this work is to explore a way to perform quantum machine learning on classical graph data. Computing a graph embedding can be very costly when the graph increases in size. There is then a hope that quantum computing will improve the speed of computing a graph embedding. Furthermore, converting classical graph data into quantum data opens new perspective for solving graph problems with quantum computers. We propose a quantum version of the Laplacian Eigenmap algorithm.

We aim to construct a quantum state whose amplitude corresponds in value to the $d$ first eigenvectors of the laplacian matrix of a graph. Our algorithm is a variational algorithm, which consist of constructing a parameterized quantum circuit, and finding the parameters that minimize a cost function. Variational circuits are very explored by the community in the past few years because they are robust to the noisy current hardware. However, there is few benchmarks that attest their utility, and even less theoretical proofs.

Some work has been done to approximate the spectral decomposition of the graph matrices by variational circuits \cite{payne2019approximate}. We go further and propose a full pipeline for solving a node classification problem with a quantum computer. We construct a graph embedding with a variational circuit, and we implement a quantum binary classifier on the top of it. We tested our algorithm on 32 nodes graphs on a quantum simulator.

This paper is the first work to the best of our knowledge to study the intersection of graph theory, machine learning and quantum computing. We put together all the pieces to enable future work on the topic. We hope that it will open new perspectives for quantum computing and quantum machine learning people to treat graph problems, and that specialists of graph machine learning and graph processing will explore quantum technologies.

We develop in the section \ref{sec:preliminaries} the tools in graph embedding theory and quantum computing theory that will serve as a basis for our work. We explain in details in section \ref{sec:qle} our main contributions, the algorithm of Quantum Laplacian Eigenmap and the quantum node classification pipeline. Section \ref{sec:experiments} details the experiments that have been realized to test our algorithm.

\label{sec:preliminaries}
\section{Preliminaries}

\subsection{Generalities on graphs}

In all this work, we will consider a graph $\mathcal{G}(V, E)$ with $V$ the set of vertices and $E$ the set of edges.  We will limit ourselves to undirected and unweighted graphs. For a vertex $i$, we will note $d_i$ its degree i.e the number of neighbors. The density of the graph $d$ is defined by the number of edges divided by the maximum number of edges, so we have $d = \frac{2|E|}{|V|(|V|+1)}$.

To any graph one can associate different matrices.
\begin{itemize}
    \item The adjacency matrix $A$ a real matrix of size $|V| \times |V|$ defined in the following way:
    \begin{equation}
        \begin{cases}
        A_{ij} = 1 &\text{if } (i,j) \in E\\
        A_{ij} = 0 &\text{otherwise}
        \end{cases}
    \end{equation}
    \item The degree matrix $D$ is the diagonal matrix such as $D_{ii} = d_i$
    \item The laplacian matrix L is of size $|V| \times |V|$ and is defined by 
    \begin{equation}
        \begin{cases}
        L_{ij} = -1 &\text{if } (i,j) \in E\\
        L_{ij} = 0 &\text{otherwise}\\
        L_{ii} = d_i 
        \end{cases}
    \end{equation}
\end{itemize}

We have the identity $L = D - A$. The Laplacian is real symmetric and therefore can be decomposed in an orthonormal eigenbase, with real eigenvalues. This property is crucial for everything that follows.

This work can be easily extended to weighted graphs, the Laplacian is still real symmetric. The generalization to directed graphs is non trivial since the previous property doesn't hold anymore. We will also assume that the number of nodes is a power of 2, and we can write $|V| = 2^n$. We can always go back to this case by adding single nodes in the graph.

\subsection{Classical Laplacian Eigenmaps}

\subsubsection{Graph embeddings}

An embedding is a map from a set of objects to an euclidean space, such as the distance in the euclidean space is representative of the similarity of the objects in the origin space. Embeddings aim to provide a vector representation of objects in order to perform analytical tasks. For machine learning tasks which take vectors as an input, such a representation is necessary. A very popular example of embeddings is word embeddings in Natural Language Processing \cite{bakarov2018survey, mikolov2013efficient, Bengio2003}. The relationship between the images of each word is representative of the semantic meaning of those words. We have for instance \textit{Paris - France + Italy = Rome} \cite{mikolov2013efficient}.

Graph embeddings can be understood two fold:
\begin{itemize}
\item Node embedding: one graph is considered, and the objective is to compute a vector representation of the nodes. One is interested by the relationship between nodes.
\item Graph embedding: a set of graph is considered and the objective is to compute a vector representation of the entire graph. One is interested to compare graphs between them.
\end{itemize}

In this paper, we will only consider node embeddings, and we will use indifferently the terms node embeddings and graph embeddings for a vector representation of nodes.

P. Goyal and E. Ferrera in \cite{Goyal2018} give an exhaustive survey of graph embedding techniques and their applications. Graph embedding algorithms can be ranged in three main categories:
\begin{itemize}
    \item Factorization methods: factorize matrices representing the connections between the nodes, such as the adjacency matrix, the laplacian matrix, or the transition probability matrix.
    \item Random walk based methods: simulate many random walks on a graph and treat them as sequences. They are very similar to NLP methods.
    \item Deep learning method: use deep autoencoders to learn the image of the nodes in a latent space. It enables to capture non-linear patterns.
\end{itemize}

The tasks for which a graph embedding is needed can be grouped in 4 categories:
\begin{itemize}
    \item Node classification: find the label of a node for which it is unobserved, predict the membership to a community. \cite{Bhagat2011}
    \item Clustering: find communities inside the graph. \cite{White2005}
    \item Link prediction: find unobserved links in the graph. \cite{Liben-Nowell2003}
    \item Visualization \cite{DiBattista1994}: architecture, hardware design.
\end{itemize}

We will note the embedding matrix of the graph $Y$, of size $2^n \times d$. $d$ is the dimension of the embedding. The row $i$, noted $Y_i$ is the vector representation of the node $i$.

\subsubsection{Laplacian Eigenmaps}

Laplacian eigenmaps \cite{Belkin2002} is a very common graph embedding and easy to compute which belongs to the factorization methods. It simply consist of taking the eigenvectors of the laplacian matrix associated with the lowest eigenvalues. The goal is to keep the first order proximity, which means that two related nodes are embedded close to each other.

The embedding $Y$ will be such as it minimizes 
$$\sum_{i,j} \|Y_i - Y_j\|^2A_{ij}$$

We have the identity 

\begin{equation}
   \sum_{i,j} \|Y_i - Y_j\|^2A_{ij} = 2 \text{tr}(Y^TLY) 
\end{equation}

The vector filled of 1s is always an eigenvector of the laplacian matrix associated to the eigenvalue 0. The eigenvalues of the Laplacian are also non-negative \cite{Spielman2012}. Therefore, a constant matrix $Y$ is a trivial solution to the above problem and it is not usable as an embedding. In order to avoid this solution, the constraint $Y^TDY = Id$ is imposed. 

The solution is given by the eigenvectors of the normalized Laplacian $D^{-1/2}LD^{-1/2}$ associated with $d$ smallest eigenvalues, counted with their multiplicity.  

In practice, one can skip the step of normalizing the laplacian matrix, and just compute the eigenvectors associated with smallest eigenvalues of $L$ \cite{VonLuxburg2007}. One just need to throw away the first eigenvector because it is the constant vector filled of 1s.

Classical ML algorithms can then be performed with the data matrix $Y$. Applying a clustering algorithm such as K-means on the top of the Laplacian Eigenmap is called spectral clustering \cite{VonLuxburg2007}.

The complexity of computing the laplacian eigenmap is $\mathcal{O}(dr|V|^2)$ with $r$ the fraction of non-zero elements in the laplacian matrix \cite{strange2014open}.

\subsection{Quantum Computing paradigms and tools}

\subsubsection{Generalities on Quantum Computing}

Quantum computing exploits the properties of quantum physics to accelerate computation \cite{M.A.NielsenandI.L.Chuang2011}. The fundamental property is that the world becomes probabilistic at small enough scale. The elementary brick of a quantum computer is the qubit. Contrary to a classical bit that has a deterministic value 0 or 1, a qubit is in a probabilistic state between 0 and 1. When it is measured, it collapses to 0 or 1. The state of a qubit is noted $|\psi\rangle = \alpha |0\rangle + \beta|1\rangle$ where $\alpha$ and $\beta$ are two complex numbers such as $|\alpha|^2 +|\beta|^2 = 1$. When measured, the qubit collapses in the state $|0\rangle$ with probability $|\alpha|^2$ or in the state $|1\rangle$ with probability $|\beta|^2$. Measuring a qubit is like flipping a coin. 

If we consider now a set of $n$ qubits, they are in a probabilistic superposition of $2^n$ states. When measured, they collapse in one of these $2^n$ states. The quantum state of the system is noted as a vector of $2^n$ entries, with the sum of square modules equal to 1.

$$|\psi\rangle = \sum_{i=0}^{2^n-1} a_i |i\rangle, a_i \in \mathbb{C}, \sum_i |a_i|^2 = 1$$

$$|\psi\rangle = 
\begin{bmatrix}
a_0\\
a_1\\
\vdots\\
a_{2^n-1}
\end{bmatrix}
$$

A quantum state is always defined up to a global phase, because a global phase is physically undetectable. It means that 
$$\sum_{k=0}^{2^n-1} a_k |k\rangle = \sum_{k=0}^{2^n-1} a_k e^{i\theta} |k\rangle \quad \forall \theta \in \mathbb{R}$$

A quantum gate is an operation that can be performed on a qubit. This operation is linear and unitary. It can be represented as a matrix $U$, its conjugate transpose is noted $U^\dagger$, and it has the property $U^\dagger U = \text{Id}$. The result of the gate on the qubit is the matrix vector product $U|\psi\rangle$. The unitary property ensures that $U|\psi\rangle$ remains a quantum state, which means it is still a normalized vector.

It exists single qubits gates and multiqubits gates. A succession of gates is a circuit. Here are a few examples of the most common gates:

\begin{itemize}
    \item H gate (Hadamard gate): $\frac{1}{\sqrt{2}}\begin{bmatrix}
    1 & 1\\
    1 & -1
    \end{bmatrix}$. Creates an uniform superposition of the qubit.
    \item X gate: $\begin{bmatrix}
    0 & 1\\
    1 & 0
    \end{bmatrix}$. Flips the qubit.
    \item CX gate: $\begin{bmatrix}
    1 & 0 & 0 & 0\\
    0 & 1 & 0 & 0\\
    0 & 0 & 0 & 1\\
    0 & 0 & 1 & 0
    \end{bmatrix}$. Flips a target qubit iif a control qubit is equal to $|1\rangle$
    \item $R_y(\theta)$ gate : $\begin{bmatrix}
    cos(\theta/2) & -sin(\theta/2)\\
    sin(\theta/2) & cos(\theta/2)
    \end{bmatrix}$. Performs a rotation around $y$ of angle $\theta$.
    \item $U(\theta, \phi, \lambda)$ gate : $\begin{bmatrix}
    cos(\theta/2) & -sin(\theta/2)e^{-i\lambda}\\
    sin(\theta/2)e^{i\phi} & cos(\theta/2)e^{i\phi+i\lambda}
    \end{bmatrix}$, $\theta \in [0,\pi], \phi, \lambda \in[0,2\pi]$. This is the general form of one qubits gate due to the unitary constraint and the invariance to a global phase.
\end{itemize}

The last two gates are parameterized gates, and play a crucial role in variational circuits design.

In a nutshell, a quantum state is a vector in a Hilbert space, which represents a probability distribution, and a quantum circuit is a linear operation on this vector.

The quantity of information in a quantum computer is exponential in the number of qubits. If one can prepare a circuit with a number of gates polynomial in the number of qubits, one can get an exponential speedup compared to a classical implementation of this task.

A number of quantum algorithms have been developed with a proven speed-up. Grover algorithm \cite{grover1996fast} for a search in an unstructured database has a quadratic speedup, Shor algorithm \cite{shor} for integer factoring has an exponential speedup, HHL \cite{hhl2009} for matrix inversion has an exponential speedup.

Such algorithms show theoretical promises, but their implementation requires millions of qubits and gates for practical problems. Today, quantum computers are in Noisy Intermediate Scale Quantum (NISQ) era. The most recent quantum computers have 50-75 qubits and can achieve about 0.1 \% error for one qubit gates and about 1\% error for 2-qubits gates \cite{pino2020demonstration}. Furthermore, there is no full connectivity between the qubits which limits the possible implementations.  

\subsubsection{Generalities on quantum machine learning}

Quantum Machine Learning is a recent and very dynamic research field which studies the intersection between quantum computing and machine learning. There are 3 approaches to combine these two disciplines, depending on the quantum nature of the model or the data \cite{schuld2018supervised}:

\begin{itemize}
    \item Classical machine learning models for quantum data. Machine Learning can help with certain tasks of quantum computing, for example discovering a quantum state with a minimal number of measurements.
    \item Quantum computing for classical data. The idea is to use quantum algorithm to perform machine learning tasks on classical data, and there is a hope that quantum algorithms can improve the performance of the current models, or can unveil new classes of models. When the expression "quantum machine learning" is used, it generally refers to this approach.
    \item Quantum machine learning models for quantum data. Quantum systems and processes will generate more and more quantum data in the future, and one may want to use machine learning to analyse these data. It is believed to me the nearest term application of quantum machine learning \cite{biamonte2017quantum} 
\end{itemize}







In this paper, our input data is a classical graph, and we are looking for an encoding of the graph into a quantum computer. This encoding can be later followed by a machine learning task. It falls then into the second, but it can also be considered in the third, if the graph features are used in combination with quantum data in a larger setting.

Let us mention 2 ways quantum computing can help with ML tasks. The first is the use of quantum linear algebra, and is a fully quantum approach. The idea is to use quantum computing to accelerate costly linear algebra tasks such as solving linear systems, finding eigenvectors and eigenvalues, and performing Fourier transformation. It is proven that algorithms like HHL \cite{hhl2009} for solving linear systems have an exponential speedup compared to their classical counterparts, but it requires a huge amount of qubits coherent for a very long time.

The second way is the use of variational algorithms to construct quantum parametric models. A quantum circuit is constructed with gates depending on parameters, and the objective is to find the parameters that minimize a cost function. These algorithms are hybrids, a quantum computer is used to compute a cost function and a classical computer updates the parameters. This is sometimes called Quantum Neural Networks \cite{broughton2020tensorflow}.

\subsubsection{Amplitude encoding}

Encoding classical data in a quantum computer is a very active research area. Different methods exist, such as basis encoding, amplitude encoding, and hamiltonian encoding. We are going to focus on amplitude encoding.

Amplitude encoding is mapping the coordinates of a unitary vector $x = (x_1, ... x_p)^T$ to the amplitudes of a quantum state. If $\|x\|=1$, and $p=2^n$, with $n$ integer, then one can naturally associate the quantum state: 
\begin{equation}
    |\psi(x)\rangle = \sum_{i=1}^{p}x_i|i\rangle
\end{equation}

Such an encoding is spacially efficient because it only needs $log(p)$ qubits to encode $x$, but it can require a circuit of depth $p$.

Let us take now a matrix $X$ of size $2^n\times 2^m$ such as $tr(XX^T)=1$, and let $X_{i\cdot}$ and $X_{\cdot j}$ respectively be the row $i$ and the column $j$. We are then able to define the quantum state of $n+m$ qubits
\begin{equation}
    |\psi(X)\rangle = \sum_{i=1}^{2^n}\sum_{j=1}^{2^m}X_{ij}|j\rangle|i\rangle
    = \sum_{i=1}^{2^n}\alpha_i|\phi_i\rangle|i\rangle = \sum_{j=1}^{2^m}\beta_j|\varphi_j\rangle|j\rangle
\end{equation}

$|\phi_i\rangle$ is the encoding of the row $i$ and $|\varphi_j\rangle$ is the encoding of the column $j$. The coefficeints $\alpha_i$ and $\beta_j$ are for normalization, and we have $\alpha_i = \|X_{i\cdot}\|$ and $\beta_j = \|X_{\cdot j}\|$. The number of required qubits varies logarithmically to the number of the coefficients of the matrix, but the required number of gates to prepare this state can be of the order of the number of coefficients.

\subsubsection{Variational Quantum Circuits}

Variational quantum circuits, or parameterized quantum circuits, are a class of circuits that gained an increasing popularity these last few years. The idea is to construct a circuit which depends to a vector of parameters $\theta$, and to minimize a cost function with reference to the parameters. More precisely, we are interested in the quantum state $|\psi(\theta)\rangle = U(\theta)|0\rangle^{\otimes n}$, with $U(\theta)$ a unitary such as $|\psi(\theta)\rangle$ represents a vector of interest. This vector is reached by finding $\theta$, which minimizes $\mathcal{C}(\theta)$. Such a setting gains interest because it is implementable with current hardware.

There are several problems in which this setting applies. $|\psi(\theta)\rangle$ can approximate a probability distribution \cite{Zoufal2019}, implement a binary classifier \cite{Schuld2018}, find the ground state of a molecule \cite{Peruzzo2014}, or approximate the solution to a combinatorial optimization problem \cite{farhi2014quantum}.

Variational quantum circuits are qualified as hybrid algorithms because the optimization part is made by a classical computer. At each step, the value of the function is computed by quantum resources, and a classical processor make the update of the parameters. Several optimization strategies can be employed, either gradient-based or not. Computing the gradients of these circuits can be a challenge, but strategies exist for particular forms of ansatz \cite{mitarai2018quantum, sweke2020stochastic}.

We will now detail 2 examples of variational circuits, the Variational Quantum Eigensolver and the Variational Quantum Classifier.

\textbf{Variational quantum eigensolver}

Variational Quantum Eigensolver (VQE) is the first variational algorithm to be introduced in the quantum computing community, and its purpose is to find the lowest eigenvalue of a hamiltonian, which was at first used to find the ground state energy of a molecule \cite{Peruzzo2014}. 

The cost function involved is the expectation of a hamiltonian $\mathcal{H}$. For any quantum state $|\psi\rangle$, $\langle \psi|H|\psi\rangle$ is lower-bounded by the smallest eigenvalue of $\mathcal{H}$, noted $\lambda_{\text{min}}(\mathcal{H})$. $\text{min}_{\theta} \langle \psi(\theta)|H|\psi(\theta)\rangle$ is then an approximation of $\lambda_{\text{min}}(\mathcal{H})$.

\textbf{Variational quantum classifier}

In \cite{Schuld2018}, Schuld at al. introduce a quantum algorithm for binary classification. We assume that a dataset $(x_1, \dots, x_N)$ is observed, with associated labels $(y_1, \dots, y_N)$. $x_i \in \mathbb{R}^{2^n}$ and $y_i \in \{0, 1\}$.

The idea is to construct a quantum circuit $|\psi(x)\rangle$ such as the probabilities of the first qubit are the probabilities of the labels. We want the first qubit to be in the state $\sqrt{\mathbb{P}(y=0)}|0\rangle + \sqrt{\mathbb{P}(y=1)}|1\rangle$.

The authors propose to construct $|\psi(x)\rangle = U(\theta) |\phi(x)\rangle$ where:
\begin{itemize}
    \item $|\phi(x)\rangle$ is an amplitude encoding of $x$, i.e $|\psi(x)\rangle = \sum_{i=0}^{2^n-1} x_i |i\rangle$, where $x_i$ is the i-th coordinate of $x$.
    \item $U(\theta)$ is a parameterized operator.
\end{itemize}

The optimization task consists to minimize a loss function over the dataset, like in a classical machine learning task.

$$ \mathcal{L}(\theta) = \sum_{i=1}^{N} l(\mathbb{P}(y_i=0|x_i;\theta), y_i)
$$

Such a formalism is identical to a binary classifier. The difference is that the probabilities $\mathbb{P}(y_i=0|x_i;\theta)$ are estimated by a quantum processor.

\subsection{Previous work}

Training quantum embeddings is a recent approach in quantum machine learning introduced by Lloyd et al. in \cite{Lloyd2020}. Instead of training a parameterized circuit to classify quantum embedded data, the author explore training the data embedding, while implementing a fixed classifier.

Concerning graph data, an attempt \cite{payne2019approximate} has been made by Payne and Srouji to use variational circuits to approximate the spectral decomposition of graphs. They propose to use the variational quantum eigensolver to estimate the eigenvalues of the graph matrices. 

\section{Quantum Laplacian Eigenmaps}
\label{sec:qle}

\subsection{General Idea}

In order for a quantum computer to perform some tasks on a graph, we need to encode the information of the graph into a quantum circuit. Classical graph embeddings aim to compute a vector representation of the nodes. In quantum node embedding, we aim to produce a quantum state which will contain the information of the graph.

There are several interests at finding a quantum embedding of the graph:
\begin{itemize}
    \item The embedding can be computed faster.
    \item Quantum computing can create different embeddings and help us capture different patterns, inaccessible with classical computers.
    \item Graph data can be associated to other quantum data in a complex process.
\end{itemize}

One way to do so can be to compute classically the node embedding matrix $\mathbf{Y}$ and to make an amplitude encoding. This solution will not give a quantum advantage, but it will create quantum data of the graph. At the end we prepare the quantum state
\begin{equation}
    |\psi\rangle = \sum_{v \in V}w_v|v\rangle |\phi_v\rangle = \sum_{v \in V}|v\rangle (Y_{v0}|0\rangle + Y_{v1}|1\rangle + ... + Y_{v,d-1}|d-1\rangle)= \sum_{v \in V}|v\rangle \left(\sum_{i=0}^{d-1}Y_{vi}|i\rangle\right)
\end{equation}

$w_v$ are weights that renormalize $|\phi_v\rangle$, ideally they are uniform.
Such a quantum state needs $log(|V|d)$ qubits to be encoded, however it needs $|V|d$ gates in the worst case for arbitrary $Y_{vi}$'s to be prepared. 

This method provides a way to encode a graph into quantum data, but doesn't solve the problem of computation time of the graph embedding, neither allow to explore new embeddings.

We propose an algorithm to construct a quantum embedding of a graph only given the laplacian matrix. It can be assimilated as a  quantum version of the classical laplacian eigenmap algorithm. Our approach is to prepare a quantum state $|\psi\rangle = \sum_{i=0}^{d-1}w_i|i\rangle|\psi_i\rangle = \begin{bmatrix}w_0|\psi_0\rangle\\ w_1|\psi_1\rangle\\ \vdots \\ w_{d-1}|\psi_{d-1}\rangle \end{bmatrix}$, with the intention that the amplitude of each $w_i|\psi_i\rangle$ encodes an eigenvector of the laplacian matrix $L$.

The idea of our algorithm is to use variational circuits to train a node embedding. Contrary to \cite{Lloyd2020}, the circuit will not just embed one node, but the whole graph. In our procedure, the embeddings will be constructed by minimizing an objective function just as in classical laplacian eigenmaps. The method looks like the one used in \cite{payne2019approximate}, but instead of measuring the eigenvalues, we use the reached quantum state as an embedding of the graph. We demonstrate that this embedding can be used for further tasks with the example of the node classification. One can also imagine applying quantum clustering algorithms such as q-means, quantum equivalent of K-means \cite{NIPS2019_8667}. To the best of our knowledge, this is the first work that proposes and implements an end to end quantum node classification algorithm.

\begin{figure}
  \centering
  \hspace*{-0.6in}
  \includegraphics[scale=0.55]{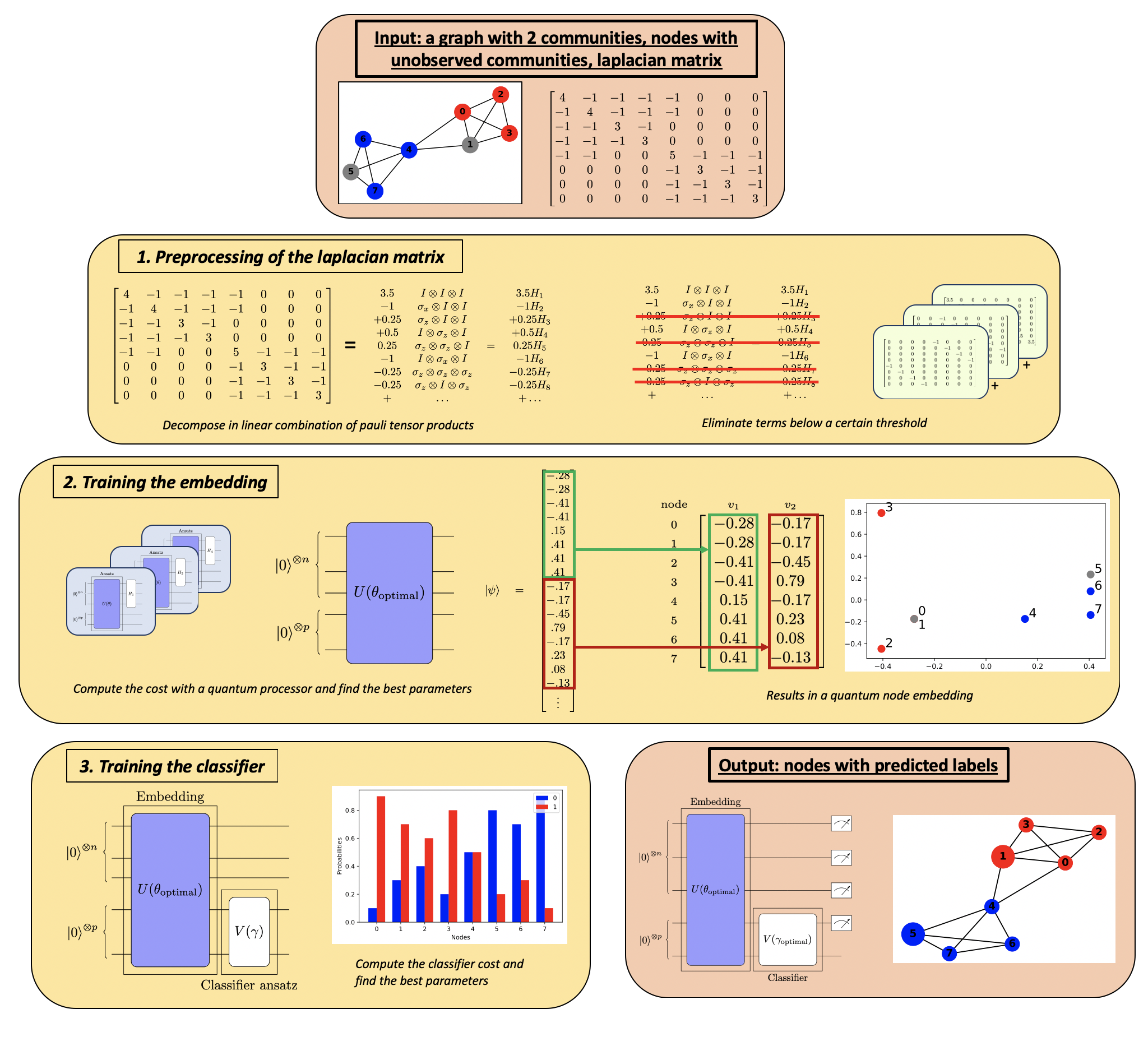}
  \caption{Overview of our algorithm.}
  \label{fig:overview}
\end{figure}

An overview of the algorithm can be seen in figure \ref{fig:overview}. The input is a graph and its laplacian matrix. As a first step, we decompose the laplacian matrix as a linear combination of pauli tensor products, and we show that we can eliminate some terms for saving computation time. This step is described in section \ref{sec:laplacian approx}. The second step described in section \ref{sec:quantum embedding} is to train the quantum embedding of the graph. At the end, we obtain a quantum state which encodes the embedding of the full graph. Finally, we show that we can train a quantum classifier on the top of the embedding to perform node classification. This is described in section \ref{sec:classification}

\subsection{Reformulation to a variational quantum circuit}

The method we propose is to use variational algorithms to produce the desired quantum state. The idea is to parameterize the state $|\psi(\theta)\rangle$ and to find the parameters that will minimize a cost function.

We are looking to the eigenvectors associated to the lowest eigenvalues of the laplacian matrix. It turns out that for a hamiltonian $\mathcal{H}$, the quantity $\langle\phi|\mathcal{H}|\phi\rangle$ has as minimum the lowest eigenvalue of $\mathcal{H}$, and the minimum is attained when $|\phi\rangle$ is the associated eigenstate. Therefore, an approximation of the lowest eigenvalue of $\mathcal{H}$ and the associated eigenvector can be computed by finding the parameters $\theta$ that minimizes the quantity $\langle\phi(\theta)|\mathcal{H}|\phi(\theta)\rangle$.

Since the Laplacian matrix is real symmetric, it can be considered as a hamiltonian, and thus finding the lowest eigenvalue and the associated eigenvector boils down to minimize $\langle\phi(\theta)|L|\phi(\theta)\rangle$ w.r.t $\theta$.

We can go further, and estimate all the $d$ first eigenvectors at the same time. Let us parameterize our state $|\psi\rangle$. We then have

$$|\psi(\theta)\rangle = \sum_{i=0}^{d-1}w_i^2|i\rangle|\psi_i(\theta)\rangle = \begin{bmatrix}w_0|\psi_0(\theta)\rangle\\ w_1|\psi_1(\theta)\rangle\\ \vdots \\ w_{d-1}|\psi_{d-1}(\theta)\rangle \end{bmatrix}$$

And we would like to minimize $\sum_{i=0}^{d-1}w_i^2\langle\psi_i(\theta)|L|\psi_i(\theta)\rangle$. If we suppose that the weights $w_i$ are all equal, and that all the $|\psi_i(\theta)\rangle$ are orthogonal, then the minimum of this quantity is reached on the first eigenvectors of $L$.

The problem is then to ensure that all the substates are orthogonal, which is no guaranteed for a general ansatz. We add then a penalty to the cost to minimize, such as the final cost function is:

$$\mathcal{C}(\theta) = \sum_{i=0}^{d-1}w_i^2\langle\psi_i(\theta)|L|\psi_i(\theta)\rangle + \lambda \sum_{i\neq j} |w_iw_j\langle\psi_i|\psi_j\rangle|^2$$

where $\lambda$ is a hyperparameter to be tuned. The cost is therefore the sum of the quantity of interest, and the constraints of orthogonality. The quadratic penalty doesn't enforce an absolute constraint of orthogonality, but avoids the system to reach an uniform superposition state. An uniform superposition state is the minimum of the first term of the cost, because it is always an eigenvector of the Laplacian with 0 as an eigenvalue, but is useless as an embedding. We don't take into account the non-uniformity of the weights $w_i$, this has to be further investigated.


Such a cost function can be estimated by a quantum circuit, and the optimisation can be made by a gradient free method such as COBYLA \cite{Powell2007}. For special forms of ansatz, the gradients of the cost function can be estimated with a quantum circuit by parameter shift rules. We don't investigate the use of gradient based optimizers in this paper, but this is certainly a way to improve our procedure even though gradients on variational quantum circuits face barren plateau problems \cite{mcclean2018barren}.

We describe in the appendix \ref{app:expectation} the quantum routine to estimate the first term of the cost function in the case where $L$ can be expressed as a tensor product of Pauli matrices. We show in the section \ref{sec:laplacian approx} that every laplacian matrix can be expressed as a linear combination of tensor product of Pauli matrices. The obtained expectations from each term of the decomposition can therefore be summed to obtain the first part of the cost function. The routines to compute the quadratic penalty is explained in appendix \ref{app:penalty} and also involves expectations of Pauli operators.

\label{sec:laplacian approx}
\subsection{The laplacian matrix as a hamiltonian}

Since the laplacian matrix is real symmetric, it can be assimilated to a quantum hamiltonian and it can be expressed as a linear combination of tensor product of Pauli matrices.

We can always write

\label{eqn:decomposition}
\begin{equation}
    L = \sum_l h_l H_l\text{ with } H_l \in U_n, \text{ and } U_n = \big\{\sigma^1_l \otimes \sigma^2_l \otimes ... \otimes \sigma^n_l \text{ with } \sigma^j_l \in \{I, X, Y, Z\}\big\}
\end{equation}

$$
I = \begin{bmatrix} 1 & 0\\ 0 & 1 \end{bmatrix}, Z = \begin{bmatrix} 1 & 0\\ 0 & -1 \end{bmatrix}, X = \begin{bmatrix} 0 & 1\\ 1 & 0 \end{bmatrix}, Y = \begin{bmatrix} 0 & -i\\ i & 0 \end{bmatrix}
$$

We prove in the appendix \ref{app:decomposition} that this decomposition always exists, is unique and can be constructed recursively. We will note $n(L)$ the number of terms in the decomposition.

\begin{figure}
  \centering
  \includegraphics[scale=0.5]{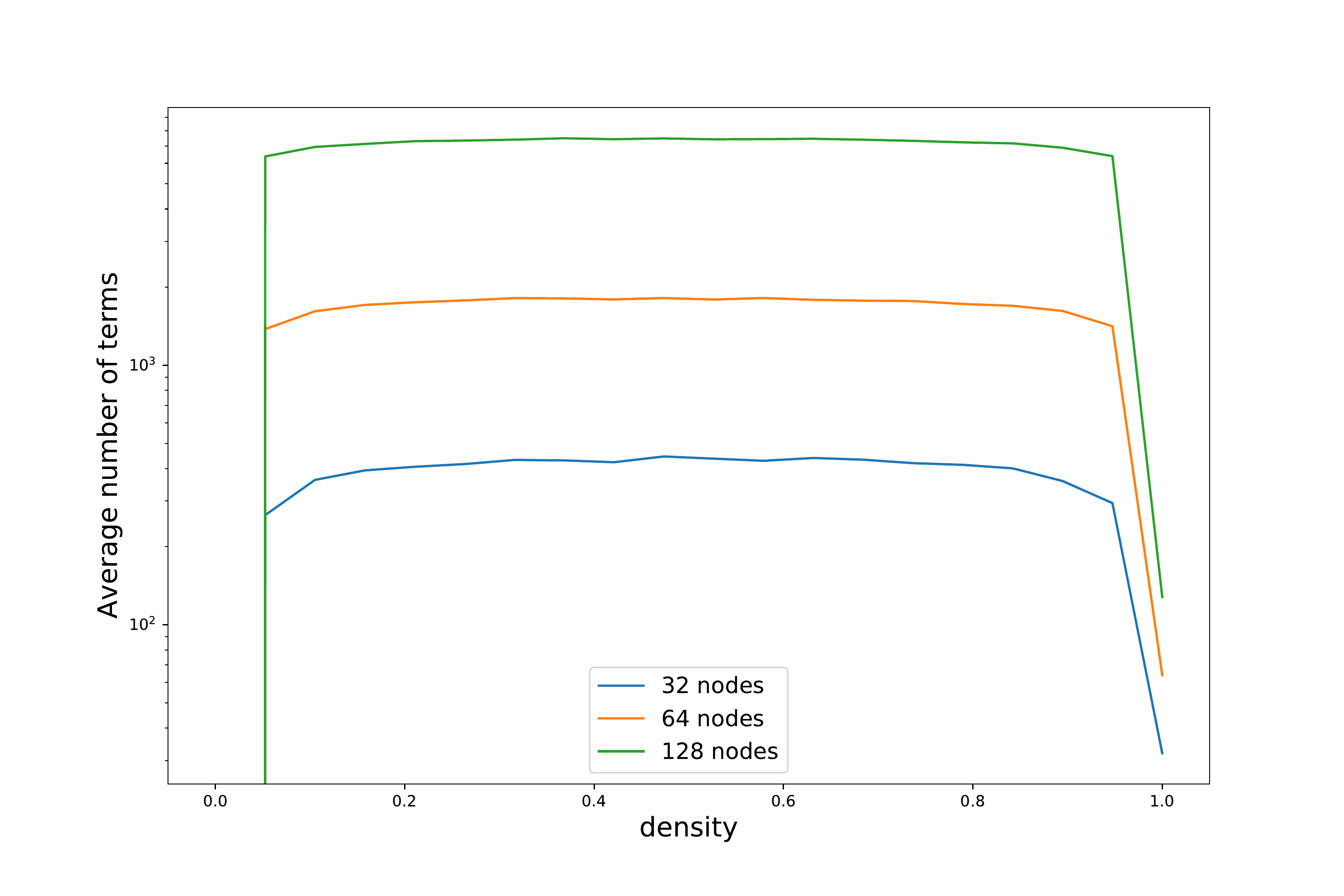}
  \caption{Number of terms in the decomposition. For each values of the density and the number of nodes, we generate 10 random graphs, and we average the number of terms in the decomposition.}
  \label{fig:nterms}
\end{figure}

The number of terms in such a decomposition can be a bottleneck for the procedure. We indeed have to execute one quantum circuit per term to estimate the expectation of the laplacian matrix. On random graphs, we observe that the number of terms is approximately equal to the number of edges. In real world problems, this number can reach several millions. Figure \ref{fig:nterms} shows the average number of terms in the decomposition for a set of 10 random graphs given the number of nodes and the density. It can be seen that this number varies almost quadratically with the number of nodes. It depends little on the density, at the exception of the extremes (fully connected and no linked graphs).

The cost of computing the expectation is linear in the number of the terms in the decomposition of the laplacian matrix thus the less terms we have to estimate the better. One solution we propose is to delete the terms for which the absolute value of $h_l$ are the lowest. In practice, we fix a threshold $t$ and we only keep the terms above the threshold. We can then write

$$
L \approx \sum_{l, |h_l|\geq t} h_l H_l
$$

It results that the quantum circuit will reach the eigenstates of the approximated laplacian matrix. This can have an effect on the performance of the algorithms, and these effects will be discussed in this work.

We define the approximation level $\alpha$ of the threshold $t$ by the following way:

\begin{equation}
    \alpha(t) = \frac{\sum_{l, |h_l|\geq t} |h_l|}{\sum_{l} |h_l|}
\label{eqn:approx}
\end{equation}

\begin{figure}
  \centering
  \includegraphics[scale=0.5]{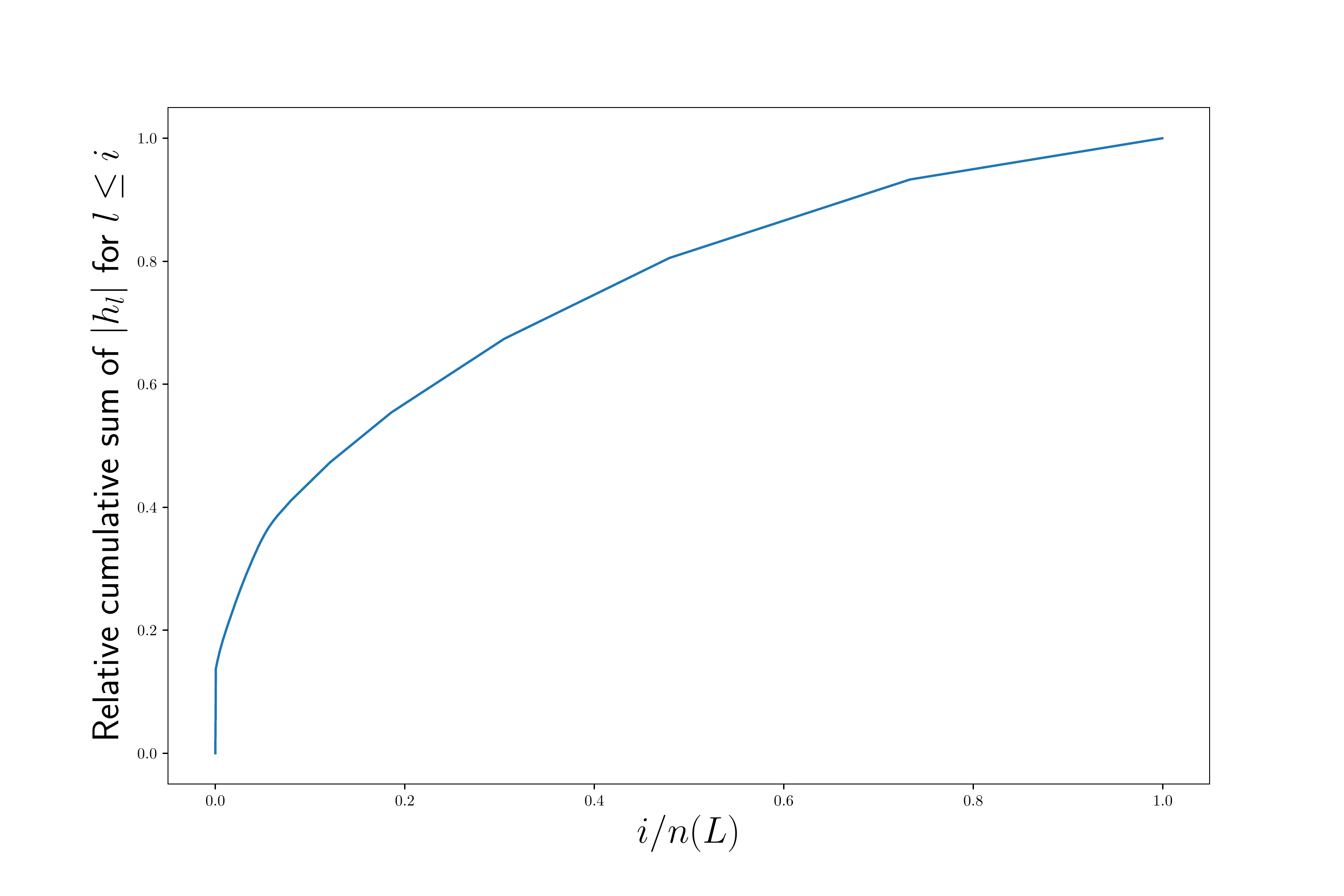}
  \caption{Relative cumulative sum of $|h_l|$ in decreasing order. The index is noted $i$. How to read: the sum of 50\% of the highest $|h_l|$ account for 80\% of the total sum. }
  \label{fig:termsrepartition}
\end{figure}

\begin{figure}
     \centering
     \begin{subfigure}[]{0.23\textwidth}
         \centering
         \includegraphics[width=\textwidth]{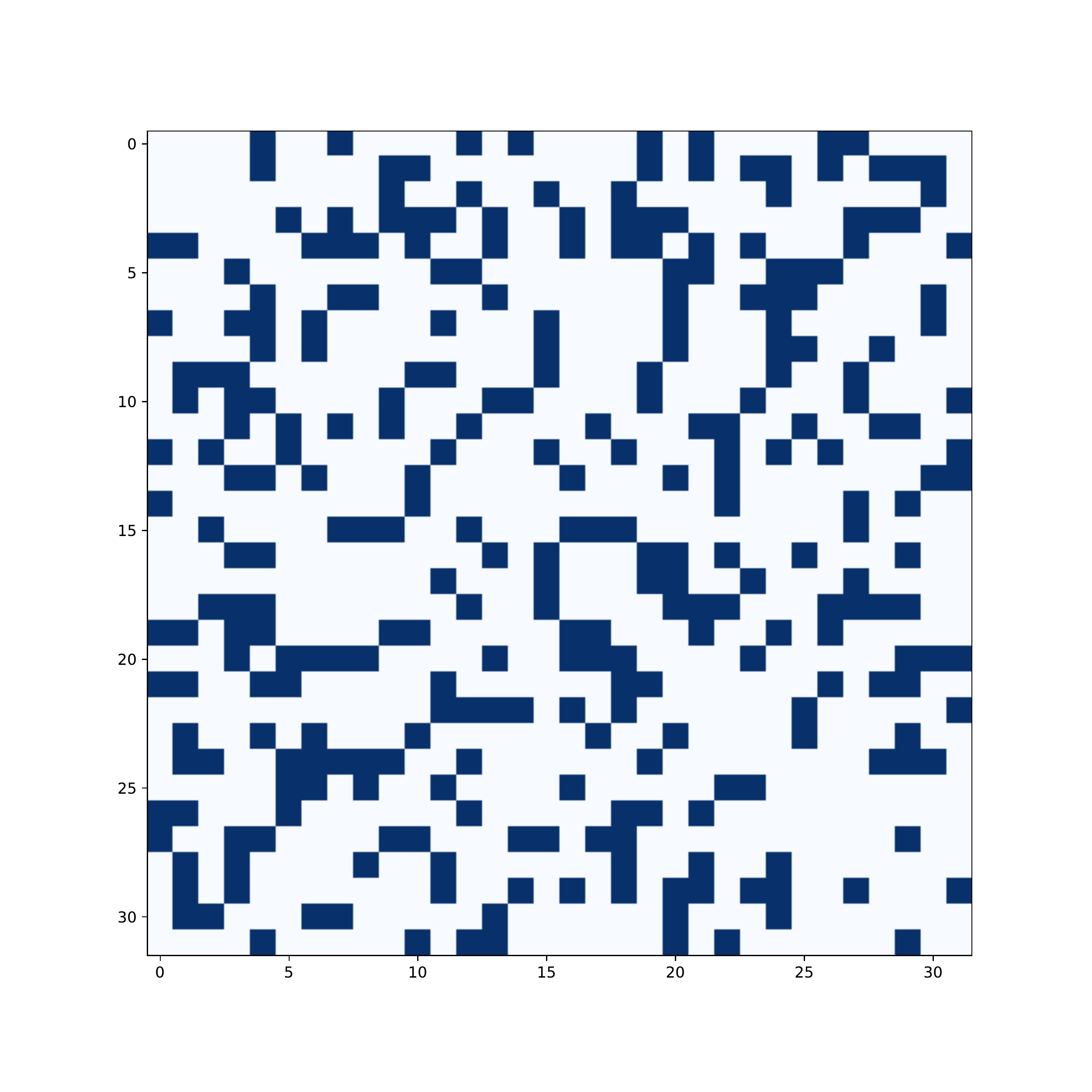}
         \caption{Adjacency matrix}
         \label{fig:y equals x}
     \end{subfigure}
     \begin{subfigure}[]{0.23\textwidth}
         \centering
         \includegraphics[width=\textwidth]{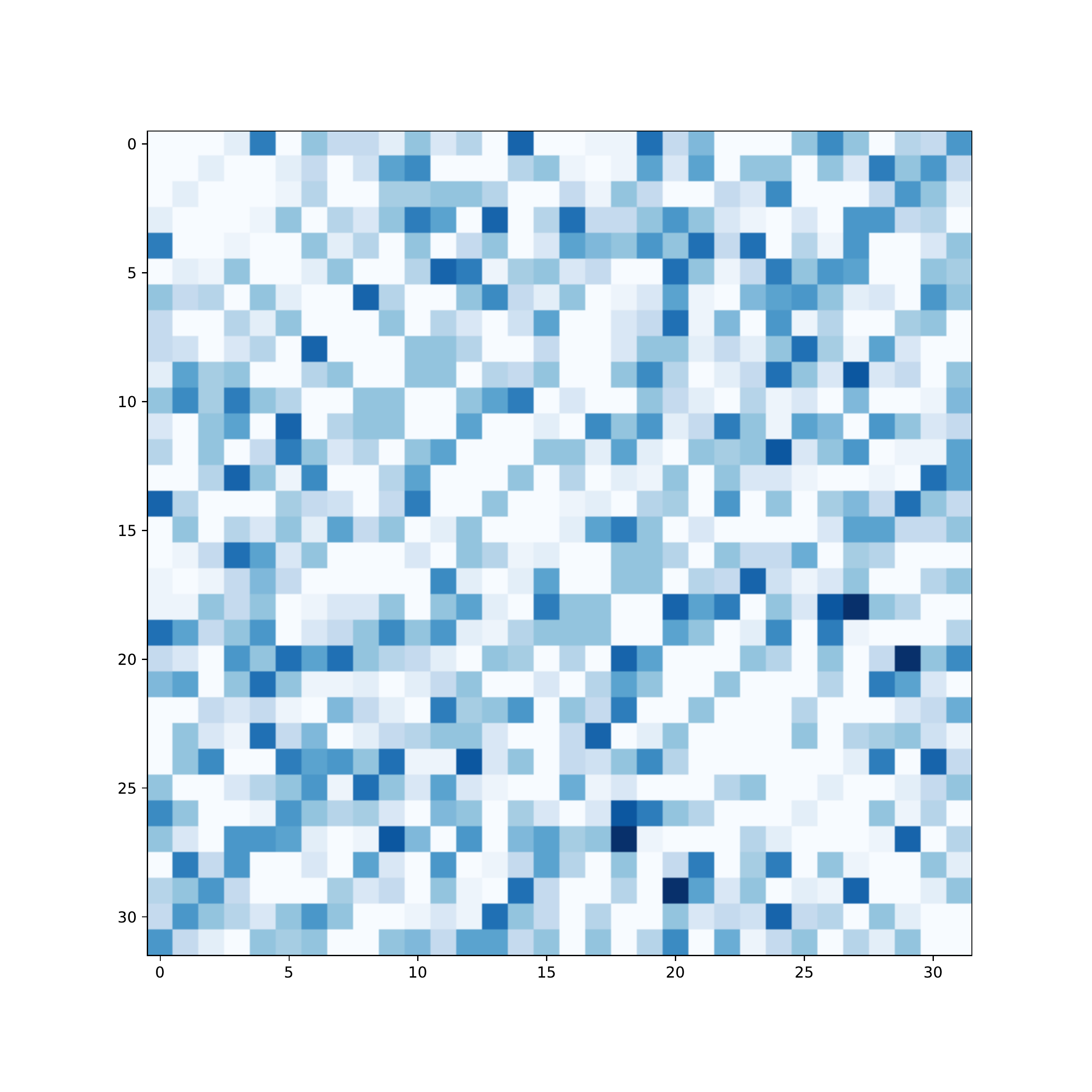}
         \caption{20\% of the terms}
         \label{fig:three sin x}
     \end{subfigure}
     \begin{subfigure}[]{0.23\textwidth}
         \centering
         \includegraphics[width=\textwidth]{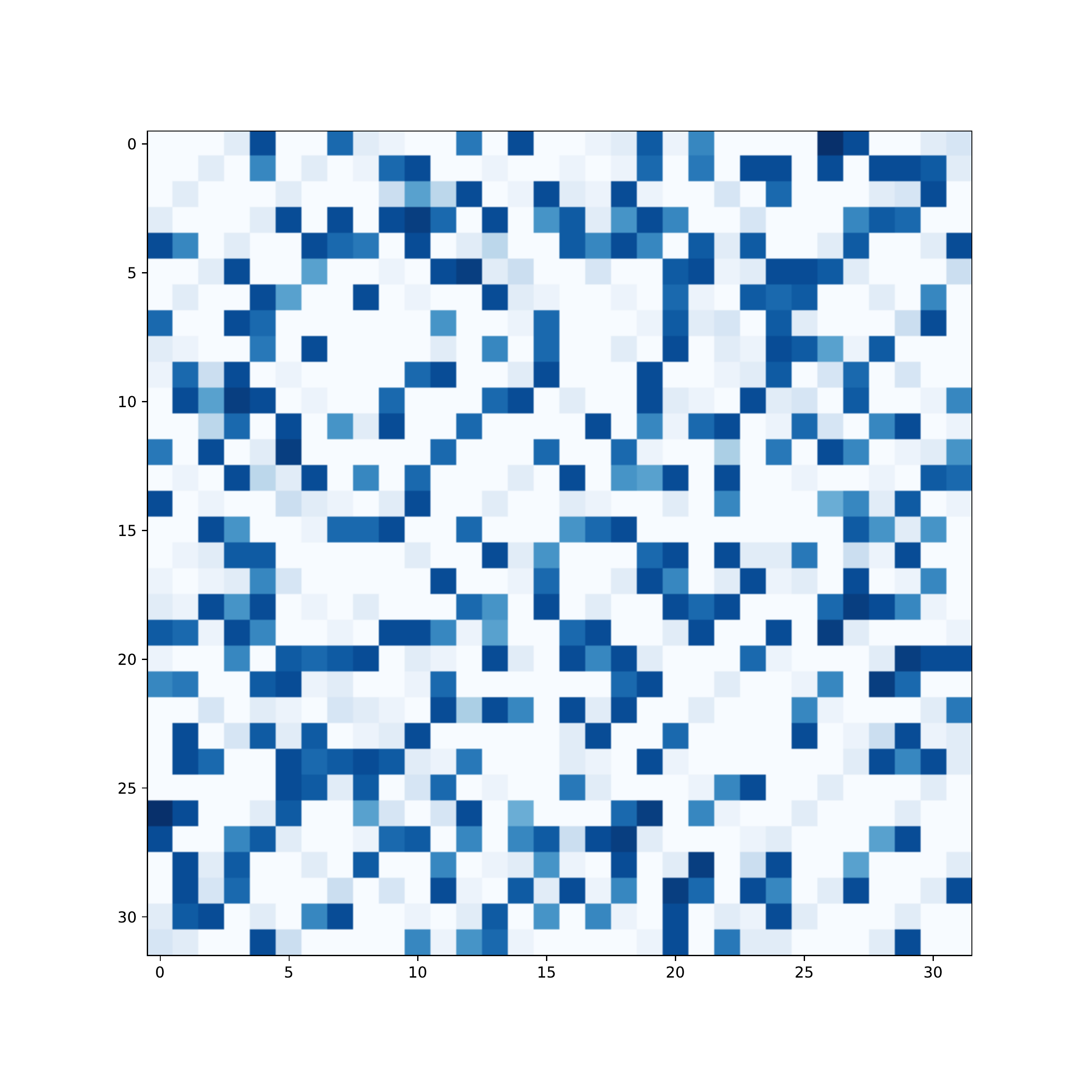}
         \caption{50\% of the terms}
         \label{fig:five over x}
     \end{subfigure}
     \begin{subfigure}[]{0.23\textwidth}
         \centering
         \includegraphics[width=\textwidth]{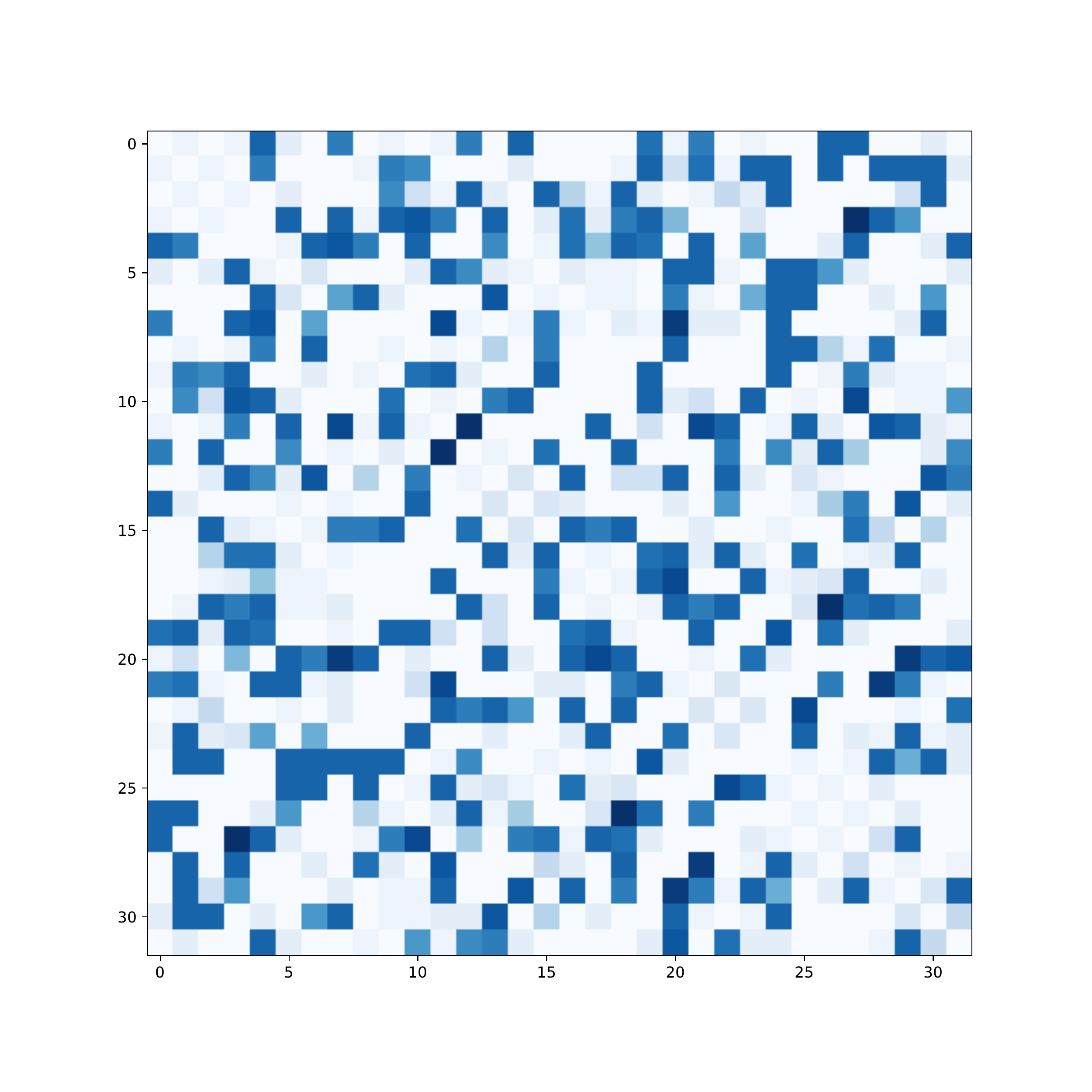}
         \caption{80\% of the terms}
         \label{fig:five over x}
     \end{subfigure}
        \caption{Visualization of the approximated matrices by keeping a top fraction of the terms in equation \ref{eqn:decomposition}}
        \label{fig:matrix reconstruction}
\end{figure}

Fortunately, we can see empirically that a few terms concentrate the majority of the objective function, so we can hope to greatly improve the computation time while not making too much error. Figure \ref{fig:termsrepartition} shows the relative cumulative sum of the ordered coefficients $|h_l|$ for a random 64 nodes graph of 0.5 density. Only one graph is displayed because this curve is almost invariant given the number of nodes and the density. It can be seen that the sum of 50\% of the highest $|h_l|$ account for 80\% of the total sum. It means that one can throw half of the decomposition and have a good approximation of the laplacian matrix. This trick can then save a lot of computational time and resources. Figure \ref{fig:matrix reconstruction} shows the approximated matrices when a top fraction of the terms are kept. It can be seen that there are few differences between the case where 50 \% of the terms are kept and the case where 80\% are kept.

Another way to reduce the computation time is to parallelize the computation on different quantum processors. The expectation of each term can indeed be evaluated independently. 

\subsection{Node Classification}

Let us suppose that each node are given a label $y_v$. The problem of node classification is to determine the label of a node given its position in the graph when its label is not observed. We will limit ourselves to a binary classification with $y_v\in \{0,1\}$. We divide the set of nodes with labels into a training test and a test set, respectively noted $V_\text{train}$ and $V_\text{test}$. 

Once we have embedded the node features in the quantum computer by the previous method, we will add a quantum classifier to the circuit. The classifier will be trained on the training set and the performances will be evaluated on the test set.

For the classification part, we suppose we have the following embedding built:
$$|\psi\rangle = \sum_{v \in V}w_v|v\rangle |\phi_v\rangle$$

We would like now to build an operator $U(\gamma)$ such as 
$$U(\gamma)|\psi\rangle = \sum_{v \in V}w_v|v\rangle U_v(\gamma)|\phi_v\rangle = \sum_{v \in V}w_v|v\rangle(a_0^v|0\rangle + a_1^v|1\rangle)|\varphi_v\rangle$$

with $|a_0^v|^2 = \mathbb{P}(y_v=0)$ and $|a_1^v|^2 = \mathbb{P}(y_v=1)$. The probabilities are estimated via measurement, and the idea is to tune $\gamma$ such as the probabilities correspond to a ground truth, which is the training set.

In practice, it only involves applying gates to $log_2(d)$ qubits, and the result can be computed by measuring $log_2(|V|)+1$ qubits.

The loss to be optimized is the binary crossentropy loss:

$$\mathcal{L}(\gamma) = \sum_{v\in V_\text{train}} y_v log(\mathbb{P}(y_v=1)) + (1-y_v) log(\mathbb{P}(y_v=0))$$

The procedure could be extended to multilabel classification by using the multiclass crossentropy as a loss function, and by measuring several qubits at the end.

\label{sec:experiments}
\section{Experiments and results}

\label{sec:setup}
\subsection{Graph configuration and setup}

\begin{figure}
  \centering
  \includegraphics[scale=0.5]{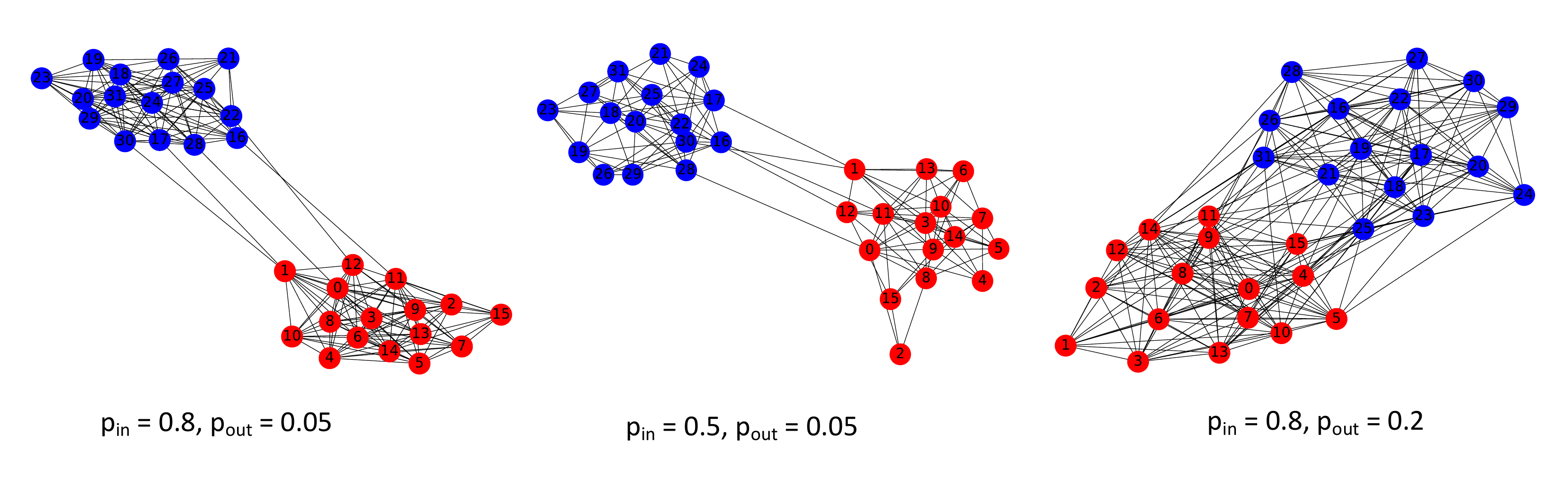}
  \caption{Examples of 32 nodes graphs.}
  \label{fig:exgraph}
\end{figure}

All our experiments will be based on a synthetic dataset of graphs that we will define. We want to create a family of graph with two communities, in order to have a pattern for classification. The graphs are parameterized by the number of nodes, the probability of connection inside a community, and the probability of connection between the communities. The parameters are respectively noted $N, p_{\text{in}}, p_{\text{out}}$. We will only consider a number of nodes which is a power of 2, so we can note $N=2^n$. Each triplet of parameters gives us several graph instances, we generate randomly different graphs and control the generated graph by the random seed. Thus, the combination $(N, p_{\text{in}}, p_{\text{out}}, \text{seed})$ characterizes an unique graph. Some examples of graphs for different values of $p_{\text{in}}$ and $p_{\text{out}}$ are shown in figure \ref{fig:exgraph}. The lower $p_{\text{in}}$ and the higher $p_{\text{out}}$, the more difficult it is to separate the two communities.

For each graph, we perform the following sequence of operations.
\begin{itemize}
    \item Split randomly the nodes in a training and a test set (25\% in the test set)
    \item Compute the tensor pauli decomposition of the laplacian matrix.
    \item Perform a thresholding operation to eliminate terms in the previous decomposition.
    \item Train the quantum embedding.
    \item Train the quantum classifier on the training set.
    \item Test the classifier on the test set, and compute the accuracy.
\end{itemize}
Both optimizations of the embedding and the classifier are done with the COBYLA algorithm \cite{Powell2007}. It is a gradient free method. To improve the quality of the solution, we launch 5 times the optimization procedure at a different random initialization point, and we keep the best. We do this for both the embedding and the classifier. We will limit ourselves to a 4 dimension embedding.

For each graph, we repeated the procedure with different values of the hyperparameter $\lambda$, and we kept the value which gives the highest accuracy on the test set. We started by the value $\lambda=100$, and we tried other values only if the accuracy test was below 0.8 because of the computational cost. If after optimization the objective function of the embedding is below the sum of the smallest eigenvalues of the Laplacian (the theoretical minimum), it means that the substates are not fully orthogonal and the penalty has to be increased, so we tested $\lambda=200$. If in the contrary it is above, the penalty may be too high, so we tested $\lambda=10$.

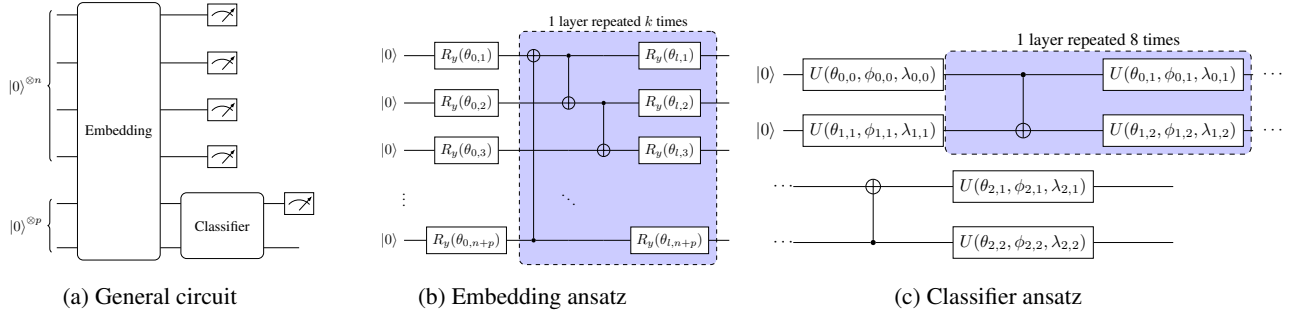
\begin{figure}
     \centering
     \begin{subfigure}[b]{0.25\textwidth}
         \centering
         \begin{tikzpicture}
        \node[scale=0.55]{
             \begin{quantikz}[thin lines]
            \lstick[wires=4]{$\ket{0}^{\otimes n}$} &\gate[wires=6, style={rounded corners}][2cm]{\text{Embedding}} & \meter{} \\
            & &\meter{} \\
            & & \meter{} \\
            & &\meter{} \\
            \lstick[wires=2]{$\ket{0}^{\otimes p}$}& & \gate[wires=2, style={rounded corners}][2cm]{\text{Classifier}}&\meter{} \\
            & & &\qw
            \end{quantikz}};
        \end{tikzpicture}
         \caption{General circuit}
         \label{fig:y equals x}
     \end{subfigure}
     \hfill
     \begin{subfigure}[b]{0.25\textwidth}
         \centering
         \begin{tikzpicture}
        \node[scale=0.55]{
             \begin{quantikz}[thin lines]
            \lstick{$\ket{0}$} & \gate{R_y(\theta_{0,1})} & \targ{}\gategroup[5,steps=4,style={dashed,
            rounded corners,fill=blue!20, inner xsep=2pt},
            background]{1 layer repeated $k$ times} & \ctrl{1}&\qw & \gate{R_y(\theta_{l,1})} & \qw\\
            \lstick{$\ket{0}$} & \gate{R_y(\theta_{0,2})} &\qw & \targ{} & \ctrl{1} & \gate{R_y(\theta_{l,2})} & \qw\\
            \lstick{$\ket{0}$} & \gate{R_y(\theta_{0,3})} &\qw & \qw& \targ{} & \gate{R_y(\theta_{l,3})} & \qw\\
            \vdots&&&\ddots&\\
            \lstick{$\ket{0}$} & \gate{R_y(\theta_{0,n+p})} & \ctrl{-4}&\qw&\qw & \gate{R_y(\theta_{l,n+p})} & \qw
            \end{quantikz}};
        \end{tikzpicture}
         \caption{Embedding ansatz}
         \label{fig:three sin x}
     \end{subfigure}
     \hfill
     \begin{subfigure}[b]{0.4\textwidth}
         \centering
         \begin{tikzpicture}
        \node[scale=0.65]{
             \begin{quantikz}[thin lines, column sep=0.2cm]
            \lstick{$\ket{0}$} & \gate{U(\theta_{0,0}, \phi_{0,0}, \lambda_{0,0})} & \ctrl{1}\gategroup[2,steps=2,style={dashed,
            rounded corners,fill=blue!20, inner sep=1pt},
            background]{1 layer repeated 8 times} & \gate{U(\theta_{0,1}, \phi_{0,1}, \lambda_{0,1})} &\qw & \cdots\\
            \lstick{$\ket{0}$} & \gate{U(\theta_{1,1}, \phi_{1,1}, \lambda_{1,1})} &\targ{} & \gate{U(\theta_{1,2}, \phi_{1,2}, \lambda_{1,2})} & \qw  & \cdots \\
            \cdots & \targ{} &\gate{U(\theta_{2,1}, \phi_{2,1}, \lambda_{2,1})} & \qw  \\
            \cdots & \ctrl{-1} & \gate{U(\theta_{2,2}, \phi_{2,2}, \lambda_{2,2})} &\qw 
            \end{quantikz}};
        \end{tikzpicture}
         \caption{Classifier ansatz}
         \label{fig:five over x}
     \end{subfigure}
        \caption{Architecture of the circuits.}
        \label{fig:circuits}
\end{figure}

The architecture of our circuits is shown in figure \ref{fig:circuits}. The circuit has in total $n+p$ qubits. The ansatz embedding is composed of one layer of parameterized $R_y$ rotations on each qubits, then $k$ layers of: 
\begin{itemize}
    \item one entangling blocks composed of a succession of CNOT gates.
    \item one layer of parameterized $R_y$ gates.
\end{itemize}
It is the same ansatz structure as in \cite{Zoufal2019}. It possesses $(k+1)\times (n+p)$ parameters. The following experiments will be limited to $k=1$ and $k=2$. The use of only $R_y$ will limit the system to reach real amplitudes. It simplifies the estimation of the penalty terms, and it is more easily comparable to the classical Laplacian Eigenmap. 

For the classifier, we adopt the same structure as in \cite{Schuld2018}, with a slightly different parameterization. The classifier ansatz only acts on 2 qubits, and it is composed of 1 layer of parameterized general unitaries followed by 8 layers of:
\begin{itemize}
    \item one  CNOT gate with alternation for each layer of the control and target qubit.
    \item one layer of parameterized general unitaries.
\end{itemize}

\subsection{General performances on 32 nodes graphs}

We evaluated the performance of our algorithm for different values of $p_{\text{in}}$ and $p_{\text{out}}$ and with different complexities of the embedding ansatz. For each values ($p_{\text{in}}$, $p_{\text{out}}$), we selected 9 seeds, and averaged the results. We performed a thresholding at 0.1, which corresponds to an approximation level of 75-85\% (the approximation level is defined in equation \ref{eqn:approx}). We compared our algorithm to the classical eigenmap algorithm combined with a logistic regression. For this purpose, we computed the 4 dimension classical eigenmap on the approximated matrix resulting of the thresholding and we trained a logistic regression on it. We also compared our algorithm to a quantum classifier trained on a random setting of the embedding parameters. We tested 2 layers ansatz and 1 layer ansatz to evaluate the effect of the complexity of the ansatz

\begin{figure}
  \centering
  \includegraphics[scale=0.5]{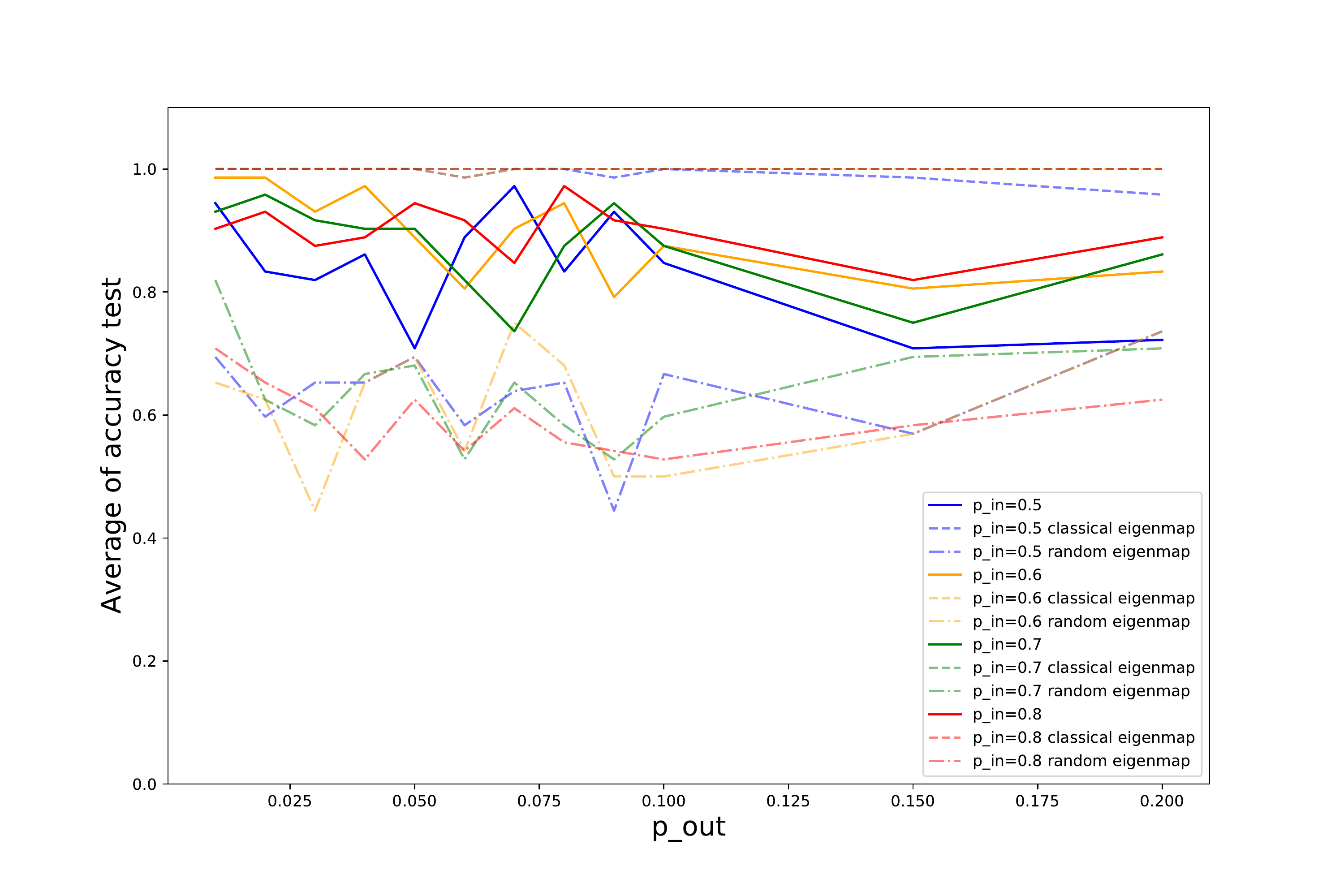}
  \caption{2 layers embedding.}
  \label{fig:result_32_2}
\end{figure}

\begin{figure}
  \centering
  \includegraphics[scale=0.5]{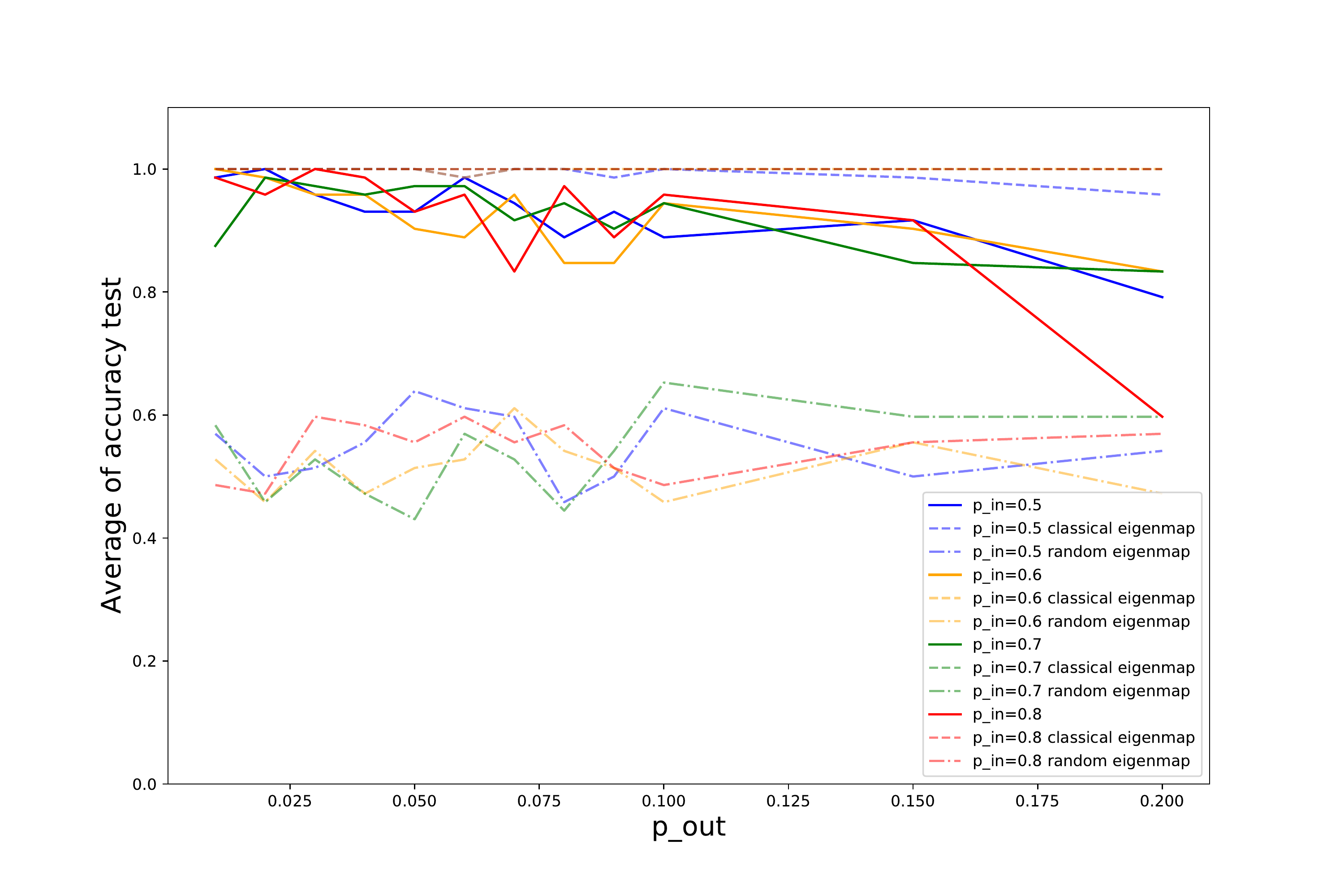}
  \caption{1 layer embedding.}
  \label{fig:result_32_1}
\end{figure}

Figure \ref{fig:result_32_2} shows the results for 2 layers embedding ansatz and figure \ref{fig:result_32_1} for 1 layer ansatz. In both cases, Our algorithm performs better than random and worse than the classical eigenmap for every value of $p_{\text{in}}$ and $p_{\text{out}}$. As expected, it performs slightly worse for higher values of $p_{\text{out}}$. Surprisingly however, 1 layer ansatz perform better than 2 layers ansatz, at least for values of $p_{\text{out}}$ below 0.1.

\subsection{Effect of thresholding}

\begin{figure}
  \centering
  \includegraphics[scale=0.5]{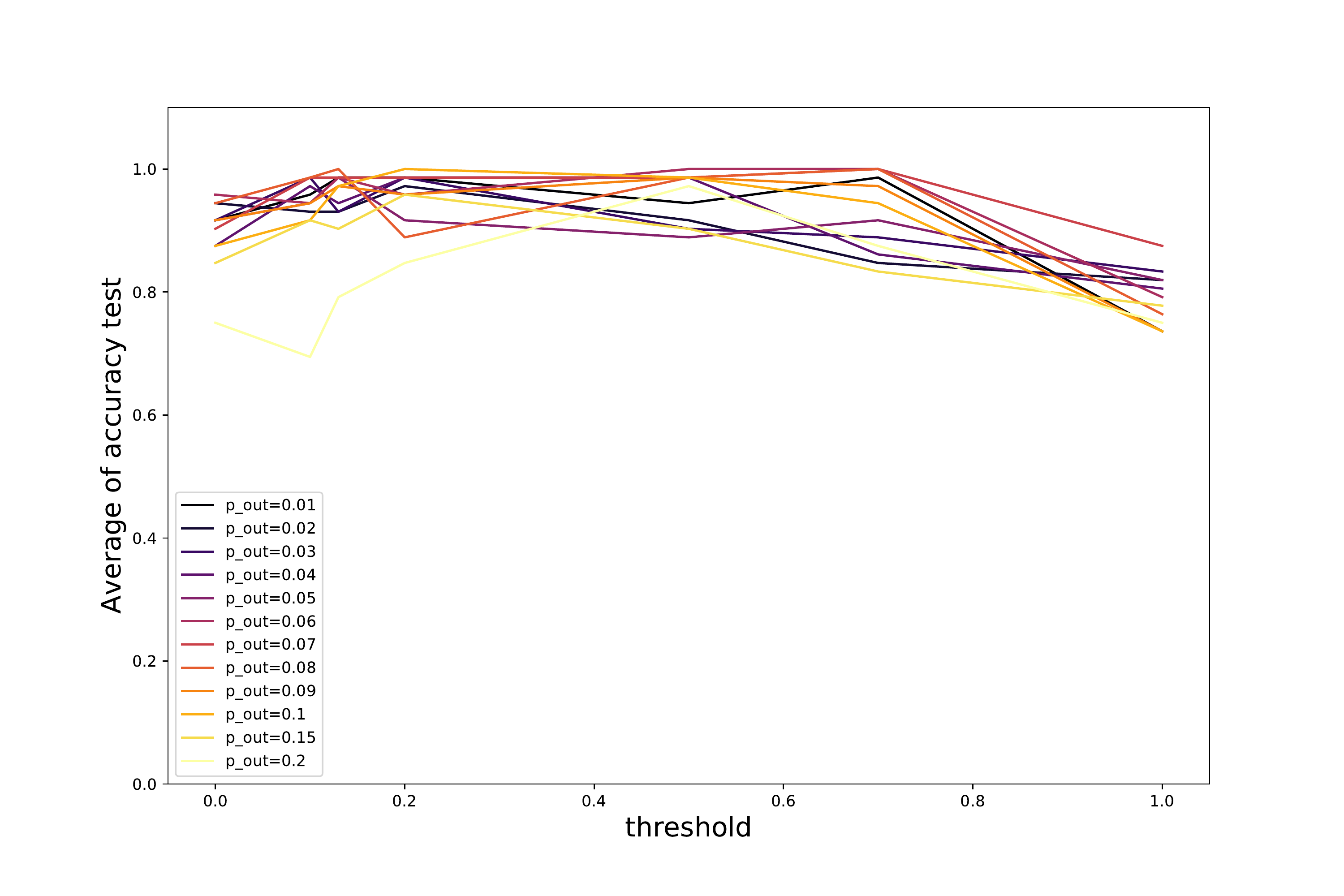}
  \caption{Effects of thresholds.}
  \label{fig:result_thres}
\end{figure}

We evaluated the effect of thresholding in the performance of the algorithm on 32 nodes graphs. We fixed the value of $p_{\text{in}}$ at 0.7, and we evaluated the algorithm for different values of $p_{\text{out}}$ and thresholds. For each couple ($p_{\text{out}}$, threshold), we selected 9 seeds and averaged the results. Figure \ref{fig:result_thres} shows the results ans table \ref{tab:approx_level} shows the approximation corresponding to the threshold levels. For a given value of threshold, the approximation level varies little for a given value of $p_{\text{out}}$ and between the different values of $p_{\text{out}}$. Therefore it is equivalent to compare threshold values and approximation level values. The experiments were made with 1 layers embeddings.

For every value of $p_{\text{out}}$, the performance of the algorithm varies little with the threshold. It implies that we can afford to cut far in the decomposition and we can save a lot in computation cost.

\begin{table}[]
    \centering
    \begin{tabular}{|c|c|c|c|c|c|c|c|}
    \hline
    \backslashbox{$p_{\text{out}}$}{threshold}  & 0 & 0.1 & 0.13 & 0.2 & 0.5 & 0.7 & 1 \\
    \hline
    0.01 & 1 & 0.84 & 0.70 & 0.59 & 0.43 & 0.32 & 0.21 \\
    0.02 & 1 & 0.84 & 0.68 & 0.57 & 0.42 & 0.31 & 0.20 \\
    0.03 & 1 & 0.82 & 0.66 & 0.55 & 0.41 & 0.31 & 0.19 \\
    0.04 & 1 & 0.80 & 0.64 & 0.54 & 0.40 & 0.29 & 0.19 \\
    0.05 & 1 & 0.79 & 0.63 & 0.53 & 0.40 & 0.29 & 0.19 \\
    0.06 & 1 & 0.80 & 0.62 & 0.52 & 0.39 & 0.29 & 0.18 \\
    0.07 & 1 & 0.80 & 0.61 & 0.52 & 0.38 & 0.29 & 0.18 \\
    0.08 & 1 & 0.79 & 0.61 & 0.50 & 0.37 & 0.28 & 0.18 \\
    0.09 & 1 & 0.82 & 0.59 & 0.51 & 0.37 & 0.28 & 0.19 \\
    0.1 & 1 & 0.79 & 0.61 & 0.50 & 0.37 & 0.27 & 0.18 \\
    0.15 & 1 & 0.81 & 0.63 & 0.50 & 0.37 & 0.27 & 0.18 \\
    0.2 & 1 & 0.85 & 0.64 & 0.52 & 0.37 & 0.28 & 0.18 \\
    \hline
    \end{tabular}
    \caption{Average of approximation level (as defined in equation \ref{eqn:approx}) for different values of $p_{\text{out}}$ and threshold}
    \label{tab:approx_level}
\end{table}

\subsection{Visualization of the quantum embeddings}

\begin{figure}
  \centering
  \includegraphics[scale=0.55]{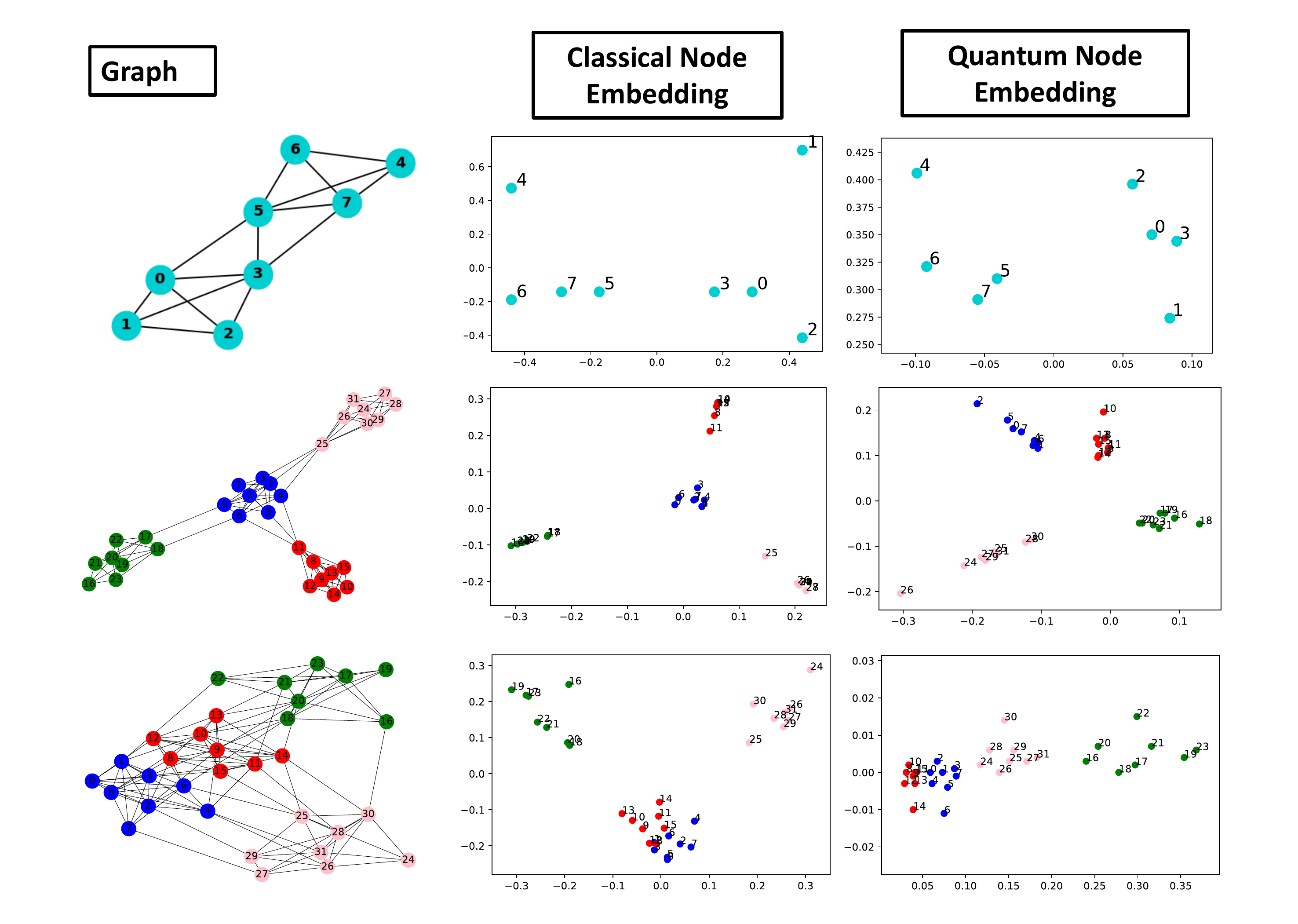}
  \caption{Visualization of the quantum embeddings compared to the classical Laplacian Eigenmap.}
  \label{fig:qual}
\end{figure}

The whole purpose in graph embeddings is to be able to capture the structure of the graph. We described in the previous subsections a quantitative way to assess the quality of our algorithm, by benchmarking the classification task on a synthetic dataset. In this part, we will visualize the quantum embeddings as we can do with classical embeddings and see if they are representative of the patterns of the graph. We propose three examples of results of our algorithm. For each example, we take a graph as an input, and we apply the procedure described in section \ref{sec:setup} until the embedding training part (we don't perform the classification). For each we plot the two first dimensions of the classical laplacian eigenmap embedding, and two dimensions of the quantum embedding which has been simulated. The results are shown in figure \ref{fig:qual}

The first example in the top is a 8 nodes graph. There is approximately 2 communities constituted by the nodes 0,1,2,3 on the one hand and the nodes 4,5,6,7 on the other hand. It can be seen that these two groups are approximately separated in the plan, both on the classical and the quantum embedding. Furthermore, the nodes 1 and 4 are separated in the graph from the rest of their community, and this is well represented in the classical embedding. Their image in the plan are apart from the rest of the nodes. We retrieve approximately the same phenomenon in the quantum embedding, more in the node 4 than in the node 1.

The second and third examples are represented in the middle and the bottom  rows of figure \ref{fig:qual}. We created 32 nodes graphs with 4 communities in different colors, very easily separable in one case, and less easily in the other one. We should expect the algorithm to embed the nodes of one community close to each other, and we should observe packed points of the same color. This is exactly what we observe in both cases. The quantum embedding separated the nodes from different communities almost as good as the classical embedding. The separation is less neat between the red and blue community in the graph of the bottom, as expected.

\section{Conclusion}

Quantum Physics offers new paradigms for handling information and provides an unique opportunity to explore new algorithms for a bunch of practical tasks. Therefore it is expected to see in the near future attempts to use quantum computing with many computational tasks.

We tried in this work to explore a way to merge graph machine learning tasks and quantum computing by constructed a quantum embedding for graph data. Graph embeddings are costly to compute, and there is a hope that quantum computing can reduce this computational time. Furthermore, it gives a way to convert graph data into quantum data on which more complex operations can be made with quantum data from other sources.

We proposed to use the amplitudes of a quantum state to store the values of a graph embedding, the laplacian eigenmap. Laplacian eigenmaps can be expressed as the minimum of a cost function. We use then variational circuits to find the parameters that minimize the cost function. At the end, the created quantum state encodes the values of the graph embedding.

The embedding can be later used for an application. We demonstrated it by building a quantum classifier circuit on the top of the graph embedding circuit. It gives us a full node classification pipeline with a quantum computer.

The algorithm was tested on different samples of graphs with communities more or less separable, and the performances are close to the classical laplacian eigenmap algorithm ones. 

We put in place all the elements required to make quantum graph machine learning algorithms. The overall procedure requires many steps, and future work has to be done to optimize each step. Each component of the algorithm can be modified, or tuned to solve a new problem. We hope then that it will open perspectives for both the communities of quantum computing and graph processing for developing new algorithms.

\appendix

\section{The decomposition of the Laplacian matrix}
\label{app:decomposition}

For any given matrix $M = \begin{bmatrix}A & B\\ C & D \end{bmatrix}$, one can always write
$$M = I\otimes\frac{A+D}{2} + Z\otimes\frac{A-D}{2} + X\otimes\frac{B+C}{2} + iY\otimes\frac{B-C}{2}$$

Applying the same scheme on the submatrices proves the existence of the decomposition.

We will now prove that this decomposition is also unique. Let us suppose that it exists two such decomposition for $L$ and that we can write 

$$
L = \sum_{H \in U_n} h_H H = \sum_{H \in U_n} h'_H H
$$

So $\sum_{H \in U_n} (h_H - h'_H) H = 0$. We have then that for all quantum state $|\psi\rangle$, $\sum_{H \in S} (h_H - h'_H) \langle\psi|H|\psi\rangle = 0$ and $\langle\psi|H|\psi\rangle$ is non zero except for a finite number of statevectors. Therefore for every $H$, $h_H - h'_H = 0$, and the decomposition is unique.

Let us now prove that $\forall l$, $h_l$ is real and $H_l$ is real.

We can write $h_l$ = $a_l +i b_l$ with $a_l$ and $b_l$ real numbers. 

We have $L = \sum_{l} a_l H_l + i \sum_{l} b_l H_l$ and for every quantum state $|\psi\rangle$,
$\langle\psi|H_l|\psi\rangle = \sum_{l} a_l \langle\psi|L|\psi\rangle + i \sum_{l} b_l \langle\psi|H_l|\psi\rangle$.

For every quantum state $|\psi\rangle$, $\langle\psi|L|\psi\rangle$ is real, and by construction, $\langle\psi|H_l|\psi\rangle$ is real for each $l$. Thus $\sum_{l} b_l \langle\psi|H_l|\psi\rangle = 0$. By the  same argument as before, every $b_l$ is null and every $h_l$ is real.

By construction of the tensor product of Pauli matrices, either $H_l$ is real or it is pure imaginary. Let $L_1 = \{ l, H_l\text{ is real}\}$ and $L_2 = \{ l, H_l\text{ is pure imaginary}\}$.

We have $L = \sum_{l \in L_1} h_l H_l + \sum_{l \in L_2} h_l H_l$, and since every $h_l$ is real and $L$ is real, we have $\sum_{l \in L_2} h_l H_l = 0$, and we can write $L = \sum_{l \in L_1} h_l H_l$. 

The decomposition is unique therefore $L_2$ is empty and every $H_l$ is real.

\section{Quantum routines}

\subsection{Compute expectations}
\label{app:expectation}

We will detail in this subsection how to compute $\langle \psi|\sigma^1_i \otimes \sigma^2_i \otimes ... \otimes \sigma^n_i|\psi\rangle = \langle \psi|\sigma|\psi\rangle$ for a given $|\psi\rangle$.

We can always decompose $|\psi\rangle$ on an eigenbase of $\sigma$, and write $|\psi\rangle = \sum_{v \in \text{eigenvectors}(\sigma)} a_v |v\rangle$ with $\sum_{v}|a_v|^2 = 1$. You then have 

$$\langle\psi|\sigma|\psi\rangle = \sum_{v} \lambda_v |a_v|^2$$

We need then to measure $|\psi \rangle$ in an eigenbasis of $\sigma$, and perform an average of the eigenvalues associated to the measured state.

However, on a standard quantum computer, we are only able to perform measurements in the computational basis. Therefore, we have to apply an operator $U$ such as $U|v\rangle = |i\rangle$. We then have

$$ U|\psi\rangle = U(\sum a_v|v\rangle) = \sum a_v U |v\rangle = \sum a_v |i\rangle$$

Measuring $|\psi\rangle$ in the eigenbasis of $\sigma$ is equivalent to measure $U|\psi\rangle$ in the computational basis.

The operator $U$ is the following:
For each qubit $j$:
\begin{itemize}
    \item Apply $H$ if $\sigma^j = X$
    \item Apply $RX(\pi/2)$ if $\sigma^j = Y$
    \item Do nothing if $\sigma^j \in \{I, Z\}$
\end{itemize}

To compute $\lambda_v$ given $|i\rangle = |q_{n-1}...q_1q_0\rangle$, one use the following expression:

\begin{equation}
    \lambda_v = \prod_j \alpha_j \text{ with } 
    \begin{cases} 
        \alpha_j = 1 & \text{if } q_j=0\\
        \alpha_j = -1 & \text{if } q_j=1  \text{ and } \sigma^j \in \{X, Y, Z\}
    \end{cases}
\end{equation}

\subsection{Compute the penalty terms}
\label{app:penalty}

We will detail in this subsection how to compute the penalty terms of the cost function. It can simply be expressed as a linear combination of expectations of pauli operators acting on the ancillas qubits.

Let's detail an example in the case of a 2-dimension embedding. We have 
$$ |\psi\rangle = w_0|0\rangle|\psi_0\rangle + w_1|1\rangle|\psi_1\rangle$$

$$\langle\psi|\sigma_x\otimes I |\psi\rangle = (w_0\langle0|\langle\psi_0| + w_1\langle1|\langle\psi_1|)(w_0|1\rangle|\psi_0\rangle + w_1|0\rangle|\psi_1\rangle) = w_0w_1\langle\psi_0|\psi_1\rangle + w_0w_1\langle\psi_1|\psi_0\rangle$$

In our case, since the ansatz are real we have $\langle\psi_0|\psi_1\rangle = \langle\psi_1|\psi_0\rangle$, and therefore

$$\langle\psi|\sigma_x\otimes I |\psi\rangle = 2w_0w_1\langle\psi_0|\psi_1\rangle$$

Elevating the last expression to square gives us the penalty term.

We will now look at the 4 dimension embedding. Every following operators will be applied to the ancillas qubits. 

$$ |\psi\rangle = w_0|00\rangle|\psi_0\rangle + w_1|01\rangle|\psi_1\rangle + w_2|10\rangle|\psi_2\rangle + w_3|11\rangle|\psi_3\rangle$$

$$\langle\psi|\sigma_x^0 |\psi\rangle = 2w_0w_1 \langle\psi_0|\psi_1\rangle + 2w_2w_3\langle\psi_2|\psi_3\rangle$$
$$\langle\psi|\sigma_z^0 |\psi\rangle = 2w_0w_1 \langle\psi_0|\psi_1\rangle - 2w_2w_3\langle\psi_2|\psi_3\rangle$$

The addition and subtraction of the two previous equations give the estimations of $w_0w_1\langle\psi_0|\psi_1\rangle$ and $w_2w_3\langle\psi_2|\psi_3\rangle$. The other terms of the penalty can be estimated on a similar way. The complexity of the procedure increases quadratically with the dimension of the embedding.











\printbibliography

@article{Goyal2018,
archivePrefix = {arXiv},
arxivId = {1705.02801},
author = {Goyal, Palash and Ferrara, Emilio},
doi = {10.1016/j.knosys.2018.03.022},
eprint = {1705.02801},
issn = {09507051},
journal = {Knowledge-Based Systems},
pages = {78--94},
title = {{Graph embedding techniques, applications, and performance: A survey}},
volume = {151},
year = {2018}
}

@article{Bhagat2011,
archivePrefix = {arXiv},
arxivId = {1101.3291},
author = {Bhagat, Smriti and Cormode, Graham and Muthukrishnan, S.},
doi = {10.1007/978-1-4419-8462-3_5},
eprint = {1101.3291},
journal = {Social Network Data Analytics},
pages = {115--148},
title = {{Node Classification in Social Networks}},
year = {2011}
}

@article{DiBattista1994,
author = {{Di Battista}, Giuseppe and Eades, Peter and Tamassiaca, Roberto and Tollis, Ioannis G},
isbn = {09257721/94},
journal = {Computational Geometry},
number = {94},
pages = {235--282},
title = {{Theory and Applications Algorithms for drawing graphs: an annotated bibliography *}},
volume = {4},
year = {1994}
}

@article{Liben-Nowell2003,
author = {Liben-Nowell, David and Kleinberg, Jon},
doi = {10.1145/956958.956972},
isbn = {1581137230},
journal = {International Conference on Information and Knowledge Management, Proceedings},
pages = {556--559},
title = {{The link prediction problem for social networks}},
year = {2003}
}

@article{White2005,
author = {White, Scott and Smyth, Padhraic},
journal = {Proceedings of the 2005 SIAM International Conference on Data Mining, SDM 2005},
pages = {274--285},
title = {{A spectral clustering approach to finding communities in graphs}},
year = {2005}
}

@article{Belkin2002,
author = {Belkin, Mikhail and Niyogi, Partlia},
doi = {10.7551/mitpress/1120.003.0080},
isbn = {0262042088},
issn = {10495258},
journal = {Advances in Neural Information Processing Systems},
title = {{Laplacian eigenmaps and spectral techniques for embedding and clustering}},
year = {2002}
}

@article{VonLuxburg2007,
archivePrefix = {arXiv},
arxivId = {0711.0189},
author = {{Von Luxburg}, Ulrike},
doi = {10.1007/s11222-007-9033-z},
eprint = {0711.0189},
issn = {09603174},
journal = {Statistics and Computing},
number = {4},
pages = {395--416},
title = {{A tutorial on spectral clustering}},
volume = {17},
year = {2007}
}

@article{Zoufal2019,
archivePrefix = {arXiv},
arxivId = {1904.00043},
author = {Zoufal, Christa and Lucchi, Aur{\'{e}}lien and Woerner, Stefan},
doi = {10.1038/s41534-019-0223-2},
eprint = {1904.00043},
issn = {20566387},
journal = {npj Quantum Information},
number = {1},
pages = {1--14},
title = {{Quantum Generative Adversarial Networks for learning and loading random distributions}},
volume = {5},
year = {2019}
}

@article{Schuld2018,
  title={Circuit-centric quantum classifiers},
  author={Schuld, Maria and Bocharov, Alex and Svore, Krysta M and Wiebe, Nathan},
  journal={Physical Review A},
  volume={101},
  number={3},
  pages={032308},
  year={2020},
  publisher={APS}
}

@article{Peruzzo2014,
archivePrefix = {arXiv},
arxivId = {arXiv:1304.3061v1},
author = {Peruzzo, Alberto and McClean, Jarrod and Shadbolt, Peter and Yung, Man Hong and Zhou, Xiao Qi and Love, Peter J. and Aspuru-Guzik, Al{\'{a}}n and O'Brien, Jeremy L.},
doi = {10.1038/ncomms5213},
eprint = {arXiv:1304.3061v1},
issn = {20411723},
journal = {Nature Communications},
number = {2},
pages = {1--10},
title = {{A variational eigenvalue solver on a quantum processor}},
volume = {5},
year = {2014}
}

@article{Lloyd2020,
archivePrefix = {arXiv},
arxivId = {2001.03622},
author = {Lloyd, Seth and Schuld, Maria and Ijaz, Aroosa and Izaac, Josh and Killoran, Nathan},
eprint = {2001.03622},
pages = {1--11},
title = {{Quantum embeddings for machine learning}},
url = {http://arxiv.org/abs/2001.03622},
year = {2020}
}

@misc{mikolov2013efficient,
    title={Efficient Estimation of Word Representations in Vector Space},
    author={Tomas Mikolov and Kai Chen and Greg Corrado and Jeffrey Dean},
    year={2013},
    eprint={1301.3781},
    archivePrefix={arXiv},
    primaryClass={cs.CL}
}

@misc{bakarov2018survey,
    title={A Survey of Word Embeddings Evaluation Methods},
    author={Amir Bakarov},
    year={2018},
    eprint={1801.09536},
    archivePrefix={arXiv},
    primaryClass={cs.CL}
}

@article{Bengio2003,
author = {Bengio, Yoshua and Ducharme, R{\'{e}}jean and Vincent, Pascal and Christian, Jauvin},
journal = {Journal of Machine Learning Research},
pages = {1137--1155},
title = {{A Neural Probabilistic Language Model}},
volume = {3},
year = {2003}
}

@book{Spielman2012,
author = {Spielman, Daniel A.},
booktitle = {Combinatorial Scientific Computing},
chapter = {18},
doi = {10.1109/focs.2007.56},
editor = {{CRC Press}},
pages = {415--525},
title = {{Spectral Graph Theory}},
year = {2012}
}

@book{strange2014open,
  title={Open Problems in Spectral Dimensionality Reduction},
  author={Strange, H. and Zwiggelaar, R.},
  isbn={9783319039435},
  lccn={2013956626},
  series={SpringerBriefs in Computer Science},
  url={https://books.google.fr/books?id=9FW6BAAAQBAJ},
  year={2014},
  publisher={Springer International Publishing}
}

@book{M.A.NielsenandI.L.Chuang2011,
author = {{M.A. Nielsen and I.L. Chuang}},
booktitle = {Contemporary Physics},
doi = {10.1080/00107514.2011.587535},
isbn = {9781107002173},
issn = {0010-7514},
number = {6},
pages = {604--605},
title = {{Quantum Computation and Quantum Information}},
volume = {52},
year = {2011}
}

@article{shor,
author = {Shor, Peter W.},
title = {Polynomial-Time Algorithms for Prime Factorization and Discrete Logarithms on a Quantum Computer},
journal = {SIAM Review},
volume = {41},
number = {2},
pages = {303-332},
year = {1999},
doi = {10.1137/S0036144598347011},
URL = {https://doi.org/10.1137/S0036144598347011},
eprint = {https://doi.org/10.1137/S0036144598347011}
}

@inproceedings{grover1996fast,
  title={A fast quantum mechanical algorithm for database search},
  author={Grover, Lov K},
  booktitle={Proceedings of the twenty-eighth annual ACM symposium on Theory of computing},
  pages={212--219},
  year={1996}
}

@article{hhl2009,
  title={Quantum algorithm for linear systems of equations},
  author={Harrow, Aram W and Hassidim, Avinatan and Lloyd, Seth},
  journal={Physical review letters},
  volume={103},
  number={15},
  pages={150502},
  year={2009},
  publisher={APS}
}

@article{Preskill2018quantumcomputingin,
  doi = {10.22331/q-2018-08-06-79},
  url = {https://doi.org/10.22331/q-2018-08-06-79},
  title = {Quantum {C}omputing in the {NISQ} era and beyond},
  author = {Preskill, John},
  journal = {{Quantum}},
  issn = {2521-327X},
  publisher = {{Verein zur F{\"{o}}rderung des Open Access Publizierens in den Quantenwissenschaften}},
  volume = {2},
  pages = {79},
  month = aug,
  year = {2018}
}

@misc{pino2020demonstration,
    title={Demonstration of the QCCD trapped-ion quantum computer architecture},
    author={J. M. Pino and J. M. Dreiling and C. Figgatt and J. P. Gaebler and S. A. Moses and M. S. Allman and C. H. Baldwin and M. Foss-Feig and D. Hayes and K. Mayer and C. Ryan-Anderson and B. Neyenhuis},
    year={2020},
    eprint={2003.01293},
    archivePrefix={arXiv},
    primaryClass={quant-ph}
}

@article{Powell2007,
author = {Powell, M. J. D.},
isbn = {DAMTP 2007/NA03},
issn = {1361-2042},
journal = {Mathematics Today-Bulletin of the Institute of {\ldots}},
number = {5},
pages = {1--12},
title = {{A view of algorithms for optimization without derivatives}},
volume = {43},
year = {2007}
}

@article{biamonte2017quantum,
  title={Quantum machine learning},
  author={Biamonte, Jacob and Wittek, Peter and Pancotti, Nicola and Rebentrost, Patrick and Wiebe, Nathan and Lloyd, Seth},
  journal={Nature},
  volume={549},
  number={7671},
  pages={195--202},
  year={2017},
  publisher={Nature Publishing Group}
}

@book{schuld2018supervised,
  title={Supervised learning with quantum computers},
  author={Schuld, Maria and Francesco Petruccione},
  year={2018},
  publisher={Springer}
}

@misc{broughton2020tensorflow,
    title={TensorFlow Quantum: A Software Framework for Quantum Machine Learning},
    author={Michael Broughton and Guillaume Verdon and Trevor McCourt and Antonio J. Martinez and Jae Hyeon Yoo and Sergei V. Isakov and Philip Massey and Murphy Yuezhen Niu and Ramin Halavati and Evan Peters and Martin Leib and Andrea Skolik and Michael Streif and David Von Dollen and Jarrod R. McClean and Sergio Boixo and Dave Bacon and Alan K. Ho and Hartmut Neven and Masoud Mohseni},
    year={2020},
    eprint={2003.02989},
    archivePrefix={arXiv},
    primaryClass={quant-ph}
}

@misc{payne2019approximate,
    title={Approximate Graph Spectral Decomposition with the Variational Quantum Eigensolver},
    author={Josh Payne and Mario Srouji},
    year={2019},
    eprint={1912.12366},
    archivePrefix={arXiv},
    primaryClass={quant-ph}
}

@article{Arute2019,
author = {Arute, Frank and Arya, Kunal and Babbush, Ryan and Bacon, Dave and Bardin, Joseph C and Barends, Rami and Biswas, Rupak and Boixo, Sergio and Brandao, Fernando G S L and Buell, David A and Burkett, Brian and Chen, Yu and Chen, Zijun and Chiaro, Ben and Collins, Roberto and Courtney, William and Dunsworth, Andrew and Farhi, Edward and Foxen, Brooks and Fowler, Austin and Gidney, Craig and Giustina, Marissa and Graff, Rob and Guerin, Keith and Habegger, Steve and Harrigan, Matthew P and Hartmann, Michael J and Ho, Alan and Hoffmann, Markus and Huang, Trent and Humble, Travis S and Isakov, Sergei V and Jeffrey, Evan and Jiang, Zhang and Kafri, Dvir and Kechedzhi, Kostyantyn and Kelly, Julian and Klimov, Paul V and Knysh, Sergey and Korotkov, Alexander and Kostritsa, Fedor and Landhuis, David and Lindmark, Mike and Lucero, Erik and Lyakh, Dmitry and Mandr{\`{a}}, Salvatore and McClean, Jarrod R and McEwen, Matthew and Megrant, Anthony and Mi, Xiao and Michielsen, Kristel and Mohseni, Masoud and Mutus, Josh and Naaman, Ofer and Neeley, Matthew and Neill, Charles and Niu, Murphy Yuezhen and Ostby, Eric and Petukhov, Andre and Platt, John C and Quintana, Chris and Rieffel, Eleanor G and Roushan, Pedram and Rubin, Nicholas C and Sank, Daniel and Satzinger, Kevin J and Smelyanskiy, Vadim and Sung, Kevin J and Trevithick, Matthew D and Vainsencher, Amit and Villalonga, Benjamin and White, Theodore and Yao, Z Jamie and Yeh, Ping and Zalcman, Adam and Neven, Hartmut and Martinis, John M},
doi = {10.1038/s41586-019-1666-5},
issn = {1476-4687},
journal = {Nature},
number = {7779},
pages = {505--510},
title = {{Quantum supremacy using a programmable superconducting processor}},
url = {https://doi.org/10.1038/s41586-019-1666-5},
volume = {574},
year = {2019}
}

@incollection{NIPS2019_8667,
title = {q-means: A quantum algorithm for unsupervised machine learning},
author = {Kerenidis, Iordanis and Landman, Jonas and Luongo, Alessandro and Prakash, Anupam},
booktitle = {Advances in Neural Information Processing Systems 32},
editor = {H. Wallach and H. Larochelle and A. Beygelzimer and F. d\textquotesingle Alch\'{e}-Buc and E. Fox and R. Garnett},
pages = {4134--4144},
year = {2019},
publisher = {Curran Associates, Inc.},
url = {http://papers.nips.cc/paper/8667-q-means-a-quantum-algorithm-for-unsupervised-machine-learning.pdf}
}

@article{commander2008,
author = {Commander, Clayton},
year = {2008},
month = {01},
pages = {},
title = {Maximum cut problem, MAX-CUT},
doi = {10.1007/978-0-387-74759-0_358}
}

@article{laporte1990selective,
  title={The selective travelling salesman problem},
  author={Laporte, Gilbert and Martello, Silvano},
  journal={Discrete applied mathematics},
  volume={26},
  number={2-3},
  pages={193--207},
  year={1990},
  publisher={Elsevier}
}

@article{mitarai2018quantum,
  title={Quantum circuit learning},
  author={Mitarai, Kosuke and Negoro, Makoto and Kitagawa, Masahiro and Fujii, Keisuke},
  journal={Physical Review A},
  volume={98},
  number={3},
  pages={032309},
  year={2018},
  publisher={APS}
}

@article{sweke2020stochastic,
  title={Stochastic gradient descent for hybrid quantum-classical optimization},
  author={Sweke, Ryan and Wilde, Frederik and Meyer, Johannes Jakob and Schuld, Maria and F{\"a}hrmann, Paul K and Meynard-Piganeau, Barth{\'e}l{\'e}my and Eisert, Jens},
  journal={Quantum},
  volume={4},
  pages={314},
  year={2020},
  publisher={Verein zur F{\"o}rderung des Open Access Publizierens in den Quantenwissenschaften}
}

@article{farhi2014quantum,
  title={A quantum approximate optimization algorithm},
  author={Farhi, Edward and Goldstone, Jeffrey and Gutmann, Sam},
  journal={arXiv preprint arXiv:1411.4028},
  year={2014}
}

@article{mcclean2018barren,
  title={Barren plateaus in quantum neural network training landscapes},
  author={McClean, Jarrod R and Boixo, Sergio and Smelyanskiy, Vadim N and Babbush, Ryan and Neven, Hartmut},
  journal={Nature communications},
  volume={9},
  number={1},
  pages={1--6},
  year={2018},
  publisher={Nature Publishing Group}
}






\end{document}